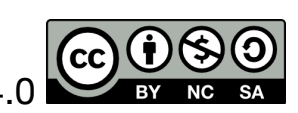

# A Quantitative Framework for Estimating System Complexity and Cost via Component Interface Analysis

Ken Y. Chan
Innovation Lab, Timeleap Inc.
Toronto, Ontario, Canada
{ken, innovation-lab}@timeleap.com

**Abstract:** *This paper introduces a formal modeling framework designed to estimate the complexity and cost associated with system changes induced by external requirements. We model a system as a directed graph of couplings, capturing the intricate dependencies and information flows between components and elements within a specific context. The proposed method enables the estimation of bounded change complexity through component interfaces, even when internal logic remains opaque. Additionally, the framework provides a mechanism for bounding the cost of system-wide modifications by associating external drivers and cost factors with individual system elements. We propose a multi-view approach to the model, providing graphical, algebraic, and tabular representations to suit different levels of abstraction and computational needs. By bridging the gap between component-based modeling and project cost estimation, our method provides actionable insights for architecture design, software engineering, and lifecycle operations. The model is validated through a case study involving the integration of a sample retail banking platform.*

## Outline

## I. Background

A fundamental tension exists in modern software engineering between the necessity for rigorous architectural decision-making and the practical constraints of project environments. While the industry increasingly relies on complex, distributed architectures, the processes used to evaluate these architectures remain largely heuristic-based. In high-pressure environments characterized by truncated timelines and budget constraints, architects often lack the necessary data to move beyond intuition. This creates a "decision-making paradox": while the cost of architectural error is high, the lack of empirical evidence forces a reliance on experience-based shortcuts. Consequently, a reliance on sub-optimal, experience-driven decisions frequently leads to significant deviations from project objectives, manifesting as cost overruns and technical debt.

[1]This material can only be used for non-commercial purposes. Please contact the author for commercial licensing terms.

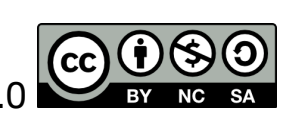

Current architectural practices leverage various frameworks (e.g., TOGAF) and design patterns (e.g., EAI, SOA, GoF) [13-15] to provide structure to the development lifecycle. Furthermore, the use of architectural patterns—ranging from high-level enterprise patterns to low-level design patterns (e.g., Singleton, Observer)—serves to standardize common solutions. However, these frameworks primarily offer structural guidance rather than evaluative capability.

While architectural patterns provide a vocabulary for design, they do not provide a mechanism for quantitative comparison. Similarly, while methodologies such as Agile or DevOps emphasize rapid iteration, they do not inherently resolve the underlying uncertainty in architectural selection. The industry lacks a unified method to bridge the gap between *applying* a pattern and *quantifying* its impact on the system's long-term viability.

The evaluation of architectural decisions is further complicated by the multi-dimensional nature of "complexity" and "cost." In software engineering, complexity is not a monolithic metric; it is a multifaceted attribute involving structural, cognitive, and operational dimensions [1-7, 20-26]. Current approaches often struggle to distinguish between architectural complexity (the arrangement of components) and implementation complexity (the intricacy of the logic).

Furthermore, the economic impact of architectural choices is difficult to forecast. Standard estimation techniques, such as those used in Waterfall project management model (e.g. work breakdown structure along with a project gantt chart) [10, 11], Agile Story Pointing [8, 12] or traditional COCOMO-based models, focus on functional throughput rather than the structural implications of architectural shifts. Because the cost of an architectural decision is often deferred to the maintenance and evolution phases of the software lifecycle, the lack of a real-time, quantitative cost-benefit analysis remains a significant barrier to optimal decision-making [9].

The literature presents a clear gap between the availability of architectural tools and the ability to perform empirical evaluations. Current computational tools (e.g. CASE tools or architectural modeling software such as Sparx Enterprise Architect [29-30], IBM Rational Rhapsody [31]) focus on the *representation* of architecture rather than the *assessment* of it. While advanced modeling techniques exist, they are often too computationally expensive or too data-intensive for use in active, fast-paced development cycles.

There is a critical need for a lightweight, mathematically grounded approach to architecture evaluation—one that can ingest structural metadata and output actionable metrics regarding complexity, risk, and estimated cost. Bridging this gap requires moving away from qualitative, experience-based assessments toward a model that can provide a quantitative, reproducible, and scalable evaluation of architectural impact.

## II. Problem and Context Definition

For most organizations, the Total Cost of Ownership (TCO)—encompassing initial development, system extensibility, and long-term maintenance—is the primary driver of project governance. However, a critical gap exists in the transparency of architectural decision-making. Stakeholders frequently struggle to reconcile business requirements with the resulting architectural complexity, specifically regarding the rationale for high-cost components and the financial implications of system interactions. There is a lack of granular, verifiable data to justify how specific architectural choices contribute to the overall cost breakdown. To address this, this study utilizes a comparative walkthrough of architectural alternatives to demonstrate the relationship between structural complexity and cost estimation.

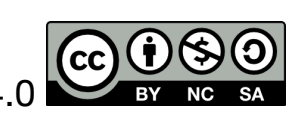

Figure 1 describes the user stories of two popular enterprise application integration (EAI) patterns: ESB [15], Service Mesh [16-18]. It shows a ESB using a centralized bus or broker (e.g. API gateway [13, 15]) to coordinate message exchanges between services. All the inbound requests from the client application to the API gateway are

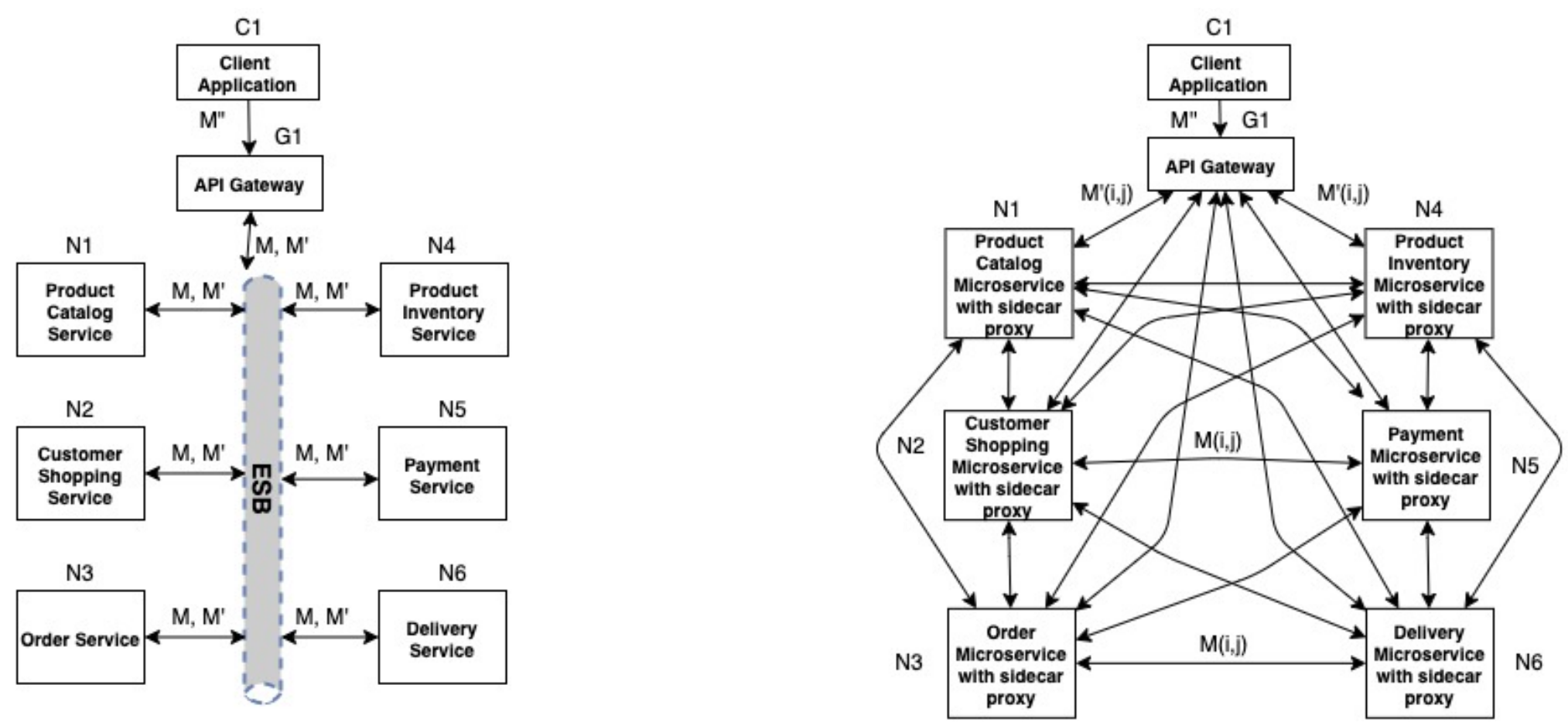

**User Stories or Use Cases:**

US1. As a shopper, I want to search the product catalog efficiently, so that I can find products of my interests to buy.

US2. As a shopper, I want to add items to my digital shopping cart, so that I do not have to checkout right away.

US3. As a shopper, I want to check out my digital shopping cart, so that I know the total cost and the availability of the items.

US4. As a shopper, I want to place an order of my checkout items, so that I can reserve the items and see the invoice.

US5. As a shopper, I want to select my delivery option for my order, so that the order will be delivered with my preferred method.

US6. As a shopper, I want to select my payment method for my order, so that I will be charged for my purchases when they have been shipped out.

**Figure 1. A Sample eCommerce System Integration - ESB vs Service Mesh Pattern**

labelled as M", while the inbound brokered messages from API gateway to the enterprise services are labelled as M'. All the internal requests from an enterprise service to another via the service broker (API gateway) are labelled as M. On the other hand, a Service Mesh may use an API gateway to broker external communications between a client application and microservices. However, the microservices communicate directly with each other via their sidecar proxies [18].

From a quick visual inspection, ESB appears to be less complex than a Service Mesh based on the number of services and the associated integration points or paths (O(N)). Additionally, Service Mesh seems to outgrow ESB in complexity with the number of integration points ($O(N+N^2)$) as the number of services (O(N)) increases. If the cost is directly proportional to the number of services (components) and the the number of integration points in a system, one would assume ESB being a better choice than Service Mesh, as in Figure 1. However, this is a simplistic view. In practice, most organizations employ two popular methods for cost-effort estimation in a project: User Story Points [8, 12] and Work Breakdown Structure (WBS) [9-11]. In the case of the user story point system, the team may use either a fixed-point (e.g. 130) or relative point system in a sprint. Each contributor then allocates a portion

[1]This material can only be used for non-commercial purposes. Please contact the author for commercial licensing terms.

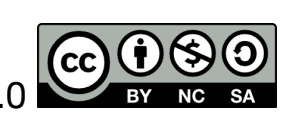

**ESB - Use Stories and Component Point Matrix**

| | US1 | US2 | US3 | US4 | US5 | US6 | Total |
|---|---|---|---|---|---|---|---|
| C1 | 1 | 1 | 1 | 1 | 1 | 1 | 6 |
| G1 | 9 | 9 | 9 | 9 | 9 | 9 | 54 |
| N1 | 10 | 0 | 0 | 0 | 0 | 0 | 10 |
| N2 | 0 | 10 | 5 | 0 | 0 | 0 | 15 |
| N3 | 0 | 0 | 0 | 10 | 0 | 0 | 10 |
| N4 | 0 | 0 | 5 | 10 | 0 | 0 | 15 |
| N5 | 0 | 0 | 0 | 0 | 5 | 5 | 10 |
| N6 | 0 | 0 | 0 | 0 | 5 | 5 | 10 |
| Total | 20 | 20 | 20 | 30 | 20 | 20 | **130** |

**Service Mesh - User Stories and Component Point Matrix**

| | US1 | US2 | US3 | US4 | US5 | US6 | Total |
|---|---|---|---|---|---|---|---|
| C1 | 1 | 1 | 1 | 1 | 1 | 1 | 6 |
| G1 | 4 | 4 | 4 | 4 | 4 | 4 | 24 |
| N1 | 20 | 0 | 0 | 0 | 0 | 0 | 20 |
| N2 | 0 | 20 | 5 | 0 | 0 | 0 | 25 |
| N3 | 0 | 0 | 0 | 20 | 0 | 0 | 20 |
| N4 | 0 | 0 | 10 | 20 | 0 | 0 | 30 |
| N5 | 0 | 0 | 0 | 0 | 10 | 10 | 20 |
| N6 | 0 | 0 | 0 | 0 | 10 | 10 | 20 |
| Total | 25 | 25 | 20 | 45 | 25 | 25 | **165** |

**ESB - Effort Estimation by Integration Point with WBS Matrix (in person-day)**

| | To: C1 | G1 | N1 | N2 | N3 | N4 | N5 | N6 | Total |
|---|---|---|---|---|---|---|---|---|---|
| From: C1 | 0 | 5 | 0 | 0 | 0 | 0 | 0 | 0 | 5 |
| G1 | 5 | 0 | 15 | 15 | 15 | 15 | 5 | 10 | 80 |
| N1 | 0 | 15 | 0 | 0 | 0 | 0 | 0 | 0 | 15 |
| N2 | 0 | 15 | 0 | 0 | 0 | 0 | 0 | 0 | 15 |
| N3 | 0 | 15 | 0 | 0 | 0 | 0 | 0 | 0 | 15 |
| N4 | 0 | 15 | 0 | 0 | 0 | 0 | 0 | 0 | 15 |
| N5 | 0 | 5 | 0 | 0 | 0 | 0 | 0 | 0 | 5 |
| N6 | 0 | 10 | 0 | 0 | 0 | 0 | 0 | 0 | 10 |
| Total | 5 | 80 | 15 | 15 | 15 | 15 | 5 | 10 | **160** |

**Service Mesh - Effort Estimation by Integration Point with WBS Matrix (in person-day)**

| | To: C1 | G1 | N1 | N2 | N3 | N4 | N5 | N6 | Total |
|---|---|---|---|---|---|---|---|---|---|
| From: C1 | 0 | 5 | 0 | 0 | 0 | 0 | 0 | 0 | 5 |
| G1 | 5 | 0 | 5 | 5 | 5 | 5 | 5 | 5 | 35 |
| N1 | 0 | 5 | 0 | 5 | 0 | 5 | 0 | 0 | 15 |
| N2 | 0 | 5 | 5 | 0 | 5 | 5 | 5 | 5 | 30 |
| N3 | 0 | 5 | 0 | 5 | 0 | 5 | 5 | 5 | 25 |
| N4 | 0 | 5 | 5 | 5 | 5 | 0 | 5 | 0 | 25 |
| N5 | 0 | 5 | 0 | 5 | 5 | 5 | 5 | 5 | 30 |
| N6 | 0 | 5 | 0 | 5 | 5 | 0 | 5 | 0 | 40 |
| Total | 5 | 35 | 15 | 30 | 25 | 25 | 30 | 40 | **205** |

**Figure 2. User Story Points vs Work Breakdown Structure - A Sample Cost Estimation for Retail Banking Integration**

of the points to the assigned user stories based on their relative priorities and complexities. In the case of WBS, the team needs to identify the tasks that may be associated with individual integration points, user stories (user cases), or components. A project manager would then consult each contributor for the effort estimate on the assigned task.

Figure 2 describes two sample User Story and Component Point Matrices for the ESB and Service Mesh patterns in Figure 1. Each table entry represents an integration point between two components. We assign user story points to entries based on their relative complexity to other components. We also assume that the communications from the client application to the API Gateway are lightweight (1 point); the communications between API gateway and services (microservices) are medium weight (external with 4 points and internal with 5 points). All user story points by service are in increments of 5 points based on the number and complexity of the interactions involved. In addition, we assume that every user story point and WBS point is translated to one person-day in effort. As a result, the total user story points in that sprint is 130 for ESB and 165 for Service Mesh, respectively.

Figure 2 also illustrates two Effort Estimation by Integration Point with WBS Matrices for ESB and Service Mesh. A table entry represents the estimated effort (in person-day) for the development task associated with the component integration point. In the case of Effort Estimation by Integration Point with WBS matrices for ESB and

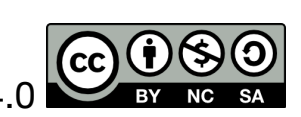

Service Mesh, we allocate in 5 effort point increments per interaction point between two services (microservices). As a result, the total amount of effort (in person-day) is 160 for ESB and 205 for Service Mesh, respectively.

Figure 3 outlines a sample enterprise architecture decision for an e-Commerce Enterprise Application Integration. It outlines the problem context for deciding whether ESB or Service Mesh should be the preferred integration pattern. Since the cost of building or maintaining the system is the key decision factor, our sample architecture decision would be primarily based on the comparison of the cost estimation matrices between ESB and Service Mesh in Figure 2. If we have a rate card of $500 USD per person-day in effort, then the labor cost for ESB and Service Mesh would be $65K USD vs $82.5K USD (130 vs 165) for User Story Points, and $80K USD vs $102.5K USD (160 vs 205) for WBS, respectively. Therefore, the decision would be to go with ESB because its complexity and cost estimates are lower than Service Mesh.

**Context, Intent, Motivations, and Applicability:**

The intent and motivation of the problem is to select an integration pattern for an eCommerce platform that allows a client application to initiate service requests to the back-end services securely and efficiently from the internet to the enterprise cloud services.

The context is for large volume transactions in e-Commerce. It is applicable to customer shopping over digital channels and also to back-end legacy system integrations.

The objectives and the driving forces are to minimize the cost of building and extending the backend services while the architecture maintains the performance requirements.

Therefore, the solution needs to keep the system integration complexity to a minimum. It must also be scalable and extendible with new applications and services while it maintains the performance requirements for the future states.

**Alternative (Structure) Description:**

In the ESB pattern (as in Figure 1), a client application would initiate service requests to the API gateway as the service broker, which routes the requests to the back-end service(s).

All the back-end services communicate with each other through the API Gateway, which acts as the service broker.

In the Service Mesh pattern (as in Figure 1), a client application would initiate service requests to the API gateway, which also routes the request to the back-end microservice(s).

All the back-end services communicate directly with each other via their sidecar proxies. Internal communications among microservices bypass the API Gateway.

All messages within a service mesh may have a unique format for every point-to-point service pair, whereas all the messages in ESB have a canonical format.

**Assumptions**

1. The cost of building and maintaining the services is the key decision factor over the performance.

2. The complexity of building a service is a key cost factor.

2. Regardless of ESB or Service Mesh, all services or microservices need to communicate and collaborate directly or indirectly with each other.

3. The cost of extending a canonical message format is much larger than a service-specific message.

**Pros for ESB**

1. Likely more complex with fewer integration points (e.g. O(N)) for future extension.

2. Likely cost less to build and manage. Message Driven and Loosely Coupled. User Story Points = 130 and WBS Effort Estimate = 160 person-days for Cost Estimation in Figure 2.

3. Easier to monitor and manage. Centralized Bus and Service Broker (API Gateway) with messages in canonical or standardized form.

**Cons for ESB**

1. Likely to cost more to build and extend the centralized bus. Still need about the same number of messages for external and internal communication via the bus (API Gateway or Broker).

2. A change to a canonical message usually results in extensive changes in Service Broker and minor changes in other services.

3. The Service Broker (API Gateway) can become monolithic and complex to extend in future.

**Pros for Service Mesh**

1. Easy to build a new integration point. Less complexity in each direct integration point.

2. Likely cost less to build. Complex Canonical Form is not required. Flexible with Direct API or Message integration via sidecar proxies.

3. Likely higher performance. Direct and decentralized communication.

**Cons for Service Mesh**

1. Harder to monitor and manage with decentralized services.

2. Likely more complex with more integration points (e.g. $O(N+N^2)$) for future extensions.

3. Likely more costly to build and extend. User Story Points = 165 and WBS Effort Estimate = 205 person-days for Cost Estimation in Figure 2.

**Decision and Rationale:**

1. Choose ESB because fewer integration points. Likely cost less to build with fewer user story points and lower WBS cost estimate.

2. Likely easier to monitor and manage while satisfying the performance objectives now.

3. Caveat: unclear on the long term outlook.

**Figure 3. A Sample Architecture Decision for eCommerce System Integration - ESB vs Service Mesh Pattern**

Traditional estimation techniques present significant challenges for long-term architectural planning. Methodologies like Story Pointing are optimized for short-term sprints, making them unsuitable for the high-level foresight required in early project phases [8]. Additionally, the lack of standardized metrics leads to inconsistent comparisons between different technical solutions. Most critically, cost estimation is often performed after a decision has already been made, rather than being used as a primary tool for decision support. While we cannot easily eliminate human subjectivity, we can address the lack of standardized, measurable tools. This work focuses on bridging that gap by providing a more consistent framework for architectural evaluation.

## III. Our Contributions

This paper presents our formal model that can estimate the complexity and cost of a system and its components solely based on the properties of the observable interfaces, with limited prior knowledge of their internals. It comes in three forms: graphical, algebraic, and tabular, which are built upon our novel definitions and theorems. It offers four key contributions that enable technologists to evaluate the complexity and cost of changes to the system due to external activities. Our model is technology independent which is especially useful in the early stage of a project because the team are still in planning mode; they do not know all the requirements and lack insights into the current state; they may have started prototyping but have not finalized many architecture decisions. Hence, the team can use our methods as a tool to refine their estimations, decisions, and scope during the course of the project.

This paper presents a novel framework for estimating the cost and complexity of software systems. The core strength of our approach is its ability to provide estimates based only on observable interface properties, even when internal implementation details are unknown. Our contributions are summarized as follows:

1. **Structural Representation via CCM and CCSM:** We introduce a method to model systems (both monolithic and distributed) using Composable Component Model (CCM) and Composable Component and System Model (CCSM). By focusing on "contextual couplings," we can represent complex interactions as a set of interconnected, observable interfaces.
2. **Predicting Complexity via SCCM:** We develop a model called System Component and Complexity Model (SCCM) in algebraic form to calculate how external changes affect a system's internal complexity. By using measurable parameters like coupling strength and interface properties, we provide a way to predict the "ripple effect" of changes without needing access to the source code.
3. **Predicting Cost via SCEM:** We extend our complexity model to include cost estimation. This allows architects to translate technical changes into budgetary impacts, providing a tool for evaluating the financial feasibility of architectural evolutions.
4. **A Scalable, Tabular Toolset:** We implement these models in a tabular format that is easy to use and scale. This allows for the rapid aggregation of data across large, complex systems, making it an effective tool for both high-level planning and detailed architectural reviews.

We demonstrate the utility of this framework through a sample retail banking system, proving that our method provides a much-needed link between software architecture and project cost management.

## IV. Modeling Principles And Assumptions

To ensure the tractability and scalability of the proposed framework, the following modeling principles are adopted:

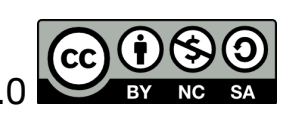

***Scope and Granularity***

The proposed model assumes a self-contained architecture where each component or system is treated as an autonomous entity [27]. To ensure model integrity, all dependencies must be explicitly defined as direct couplings; we exclude implicit dependencies and shared components between systems [4]. While we acknowledge that indirect interactions and secondary side effects contribute to total system complexity [33-34], this study focuses exclusively on primary-order, explicitly mapped couplings to ensure computational tractability.

Our model adopts an interface-centric approach, focusing on the observable interactions between components rather than their internal logic. Traditionally, coupling and cohesion [5-7] as well as entropy [26] are important concepts in complexity. Nonetheless, we assume that the complexity of a system is primarily driven by its structural connections—specifically, the "contextual" interactions between elements. Consequently, we focus on **contextual couplings** rather than internal attributes like cohesion. While we acknowledge that internal complexities (e.g., concurrency [3], algorithmic complexity [21-22], or side effects [33-34]) exist, we assume they are secondary to the structural overhead of managing inter-component dependencies. Furthermore, the model focuses on the boundaries of the system; it assumes that while a system may rely on external services, the estimation is bounded by the visible interfaces (APIs) of the components in scope.

***Structural and Mathematical Foundations***

The framework is built upon graph-theoretic foundations, where the system is represented as a set of vertices (V) and edges (E). We assume that:

- **Independence of Connections:** Each edge represents a discrete interaction, and the impact of one connection is mathematically independent of others within the same layer.
- **Attribute Attribution:** Each vertex and edge possesses measurable, quantitative attributes (e.g., throughput, frequency, or error rate) that can be used to drive cost/complexity estimations.
- **Upper-Bound Estimation:** To account for uncertainty in unobserved internal logic, the model adopts a conservative estimation strategy, calculating the upper bound of complexity based on the potential interaction of all identified interfaces.

***Drivers of Change and Complexity***

The model assumes that changes to the system are exogenous, driven by external actors performing specific activities within a defined context. We assume that the primary driver of complexity is the requirement to adapt existing interfaces to accommodate new operational needs. While we do not model the fine-grained evolution of code, we assume that the structural impact of these changes—namely the addition or modification of connection parameters—is a sufficient proxy for the effort required to maintain the system.

## V. Graphical Models for Composable Components and System (CCM, CCSM)

Complexity is multifaceted and domain-specific [23]; however, this work addresses 'organized complexity'—the web of interactions between system elements. Drawing on the principles of component-based architecture [32], we define components as functional entities characterized by their interfaces. This allows for a bounded cost estimation approach that relies solely on quantifiable, external attributes. By bypassing the need for visibility into internal component logic, our model provides a scalable way to estimate system costs based on observable interface properties.

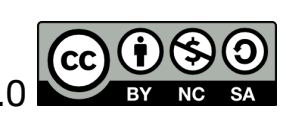

The system architecture is formally modelled using graph theory, where components are represented as nodes (V) and interactions or couplings as edges (E). The graph structure (G = (V, E)) inherently captures the structural dependencies [19]. Crucially, this graph is not treated as purely static; its nodes and edges are enriched by a multi-dimensional contextual layer. This context allows the model to extend beyond mere connectivity, capturing critical metadata along several axes: the Domain Context (e.g., software, process, or business domain); the Operational Context (e.g., the deployment environment or lifecycle stage); and the Governance Context (e.g., security, compliance, or ownership). By attaching these contextual attributes to the graph's elements, we enable complexity analysis that accounts for the system's operational environment, lifecycle phase, and regulatory constraints, thereby elevating the model from a simple network diagram to a rich, context-aware knowledge graph.

**Figure 4. A Basic Composable Component Model (CCM)**

**Figure 4a. Component Interfaces and Implementation**

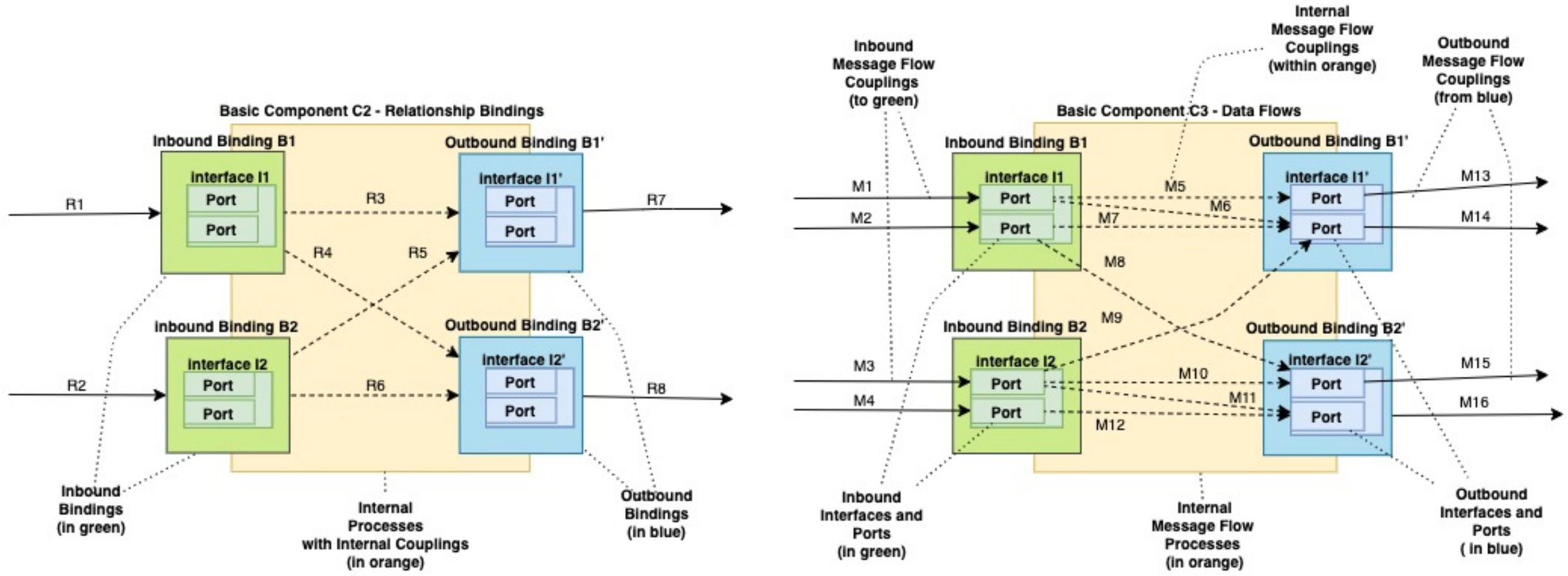

**Figure 4b. Internal and External Couplings with Relationship Bindings**

**Figure 4c. Internal and External Couplings with Data Flows**

To formalize the structural and behavioural dynamics of system units, we introduce the Composable Component Model (CCM). This model represents a component as a highly structured graph, allowing for both the visualization and quantitative analysis of its internal and external dependencies, as illustrated in Figures 4. A Component is defined as a composable, self-contained system unit. Its compositionally stems from the fact that it exposes its properties and functions through fundamental Elements. An Element is defined as an indivisible, atomic unit that serves as a mandatory interaction point. These elements include abstract concepts like interfaces, concrete services, ports, bindings, and feature sets. An element's role is not only to represent a type or an instance of an entity but also to function as a primary connector within the overall graph.

[1]This material can only be used for non-commercial purposes. Please contact the author for commercial licensing terms.

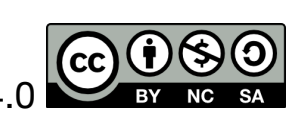

A coupling formally represents the coupling strength between two interacting elements [6]. We classify these couplings into two primary modalities:

- *A re*lational coupling represents logical dependencies or informational dependencies between elements; for example, a relationship, association, aggregation, or dependency from one element to another element.
- A temporal (or data-flow) coupling represents dependencies based on sequence or timing; for example, a data (message) flow from one element to another element.

The overall system complexity is defined by the cumulative strength of these couplings. The framework mandates that components are viewed as heterogeneous collections of elements, unified by these defined coupling structures.

When examining a basic component, the model uses boundaries (defined by internal/external interfaces and ports) to rigorously separate its internal logic from its external contracts, as illustrated in Figure 4. A component is connected through specific couplings, which are formally tagged using prefixes to denote their nature: D (Dependency), R (Relationship), or M (Message flow). These prefixes map directly to parameterized object types defining the nature of the connection. We distinguish two sets of couplings:

- Internal Couplings: Reside entirely within the component boundary (often visualized as dotted arrows).
- External Couplings: Reside outside the component boundary (often visualized as solid arrows).

Crucially, while external coupling is typically observed directly through the exposed interfaces, determining the total internal coupling strength can be difficult. Therefore, we propose an estimation mechanism: the upper bound (maximum) internal coupling amount is calculated by evaluating the maximum potential connections between a component's defined inbound and outbound interfaces.

In applied terms, the CCM is analogous to Service-Oriented Architecture (SOA). A component can also be conceptualized as a microservice, featuring inbound service interfaces (APIs) and an internal service implementation. The inbound APIs serve to aggregate and normalize initial inputs, while the internal processes generate and distribute outputs to the outbound interfaces [32]. These outbound interfaces then route the outputs to target components, completing the flow of service execution [15–17, 32]. This framework provides a robust means of analyzing service cohesion and defining granular separation of concerns.

Figure 4a shows a basic component C1 in our CCM that has two types of sub-components: *client interface (API) and component implementation* [32]. At the core of the architecture, a component is defined by a set of explicit interfaces. These interfaces encapsulate the service contract and define the boundaries of interaction. The functional exposure points of these interfaces are realized as ports. These ports are not merely endpoints; they are type-constrained gateways that mandate the specific types of data or state changes permitted between components.

The interaction points are further refined by binding mechanisms. For instance, an interface port defines the required inputs and outputs, constraining data integrity and ensuring type safety across component boundaries. This structural discipline guarantees that communication failures can be accurately localized to boundary violations rather than internal logic errors.

Figure 4b, illustrated by Component C2 describes Relational/Binding Coupling. This architecture defines a component structure comprising distinct inbound and outbound bindings, which function as the formalized

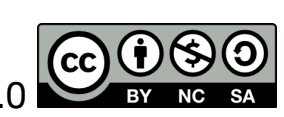

endpoints for all external component interactions. C2 establishes a coupling by projecting its outbound binding contract onto the inbound binding contract of the target component. Internally, the processes within C2 mediate the flow by establishing internal couplings that connect inbound dependencies to subsequent outbound dependencies. Analogously, these internal processes execute a critical mixing function, wherein the initial consumed inputs are transformed and logically connected to the resultant outputs, indicating that the component's functionality is highly dependent on the shared, evolving state defined by these bindings.

In contrast, Figure 4c, illustrated by Component C3, utilizes Data-Flow/Message Coupling. This mechanism replaces the dependency on shared state with a focus on passing immutable data payloads. Component C3 is characterized by inbound and outbound interface ports, each of which must be strictly associated with a singular message type. Data-flow coupling operates by transmitting a message from an outbound interface port to an inbound interface port via a dedicated message channel. Consequently, the interaction is entirely payload-driven. Furthermore, the internal processes within C3 execute a fan-in and fan-out function, consuming specific inputs from inbound interfaces to produce and distribute structured outputs through the outbound interfaces.

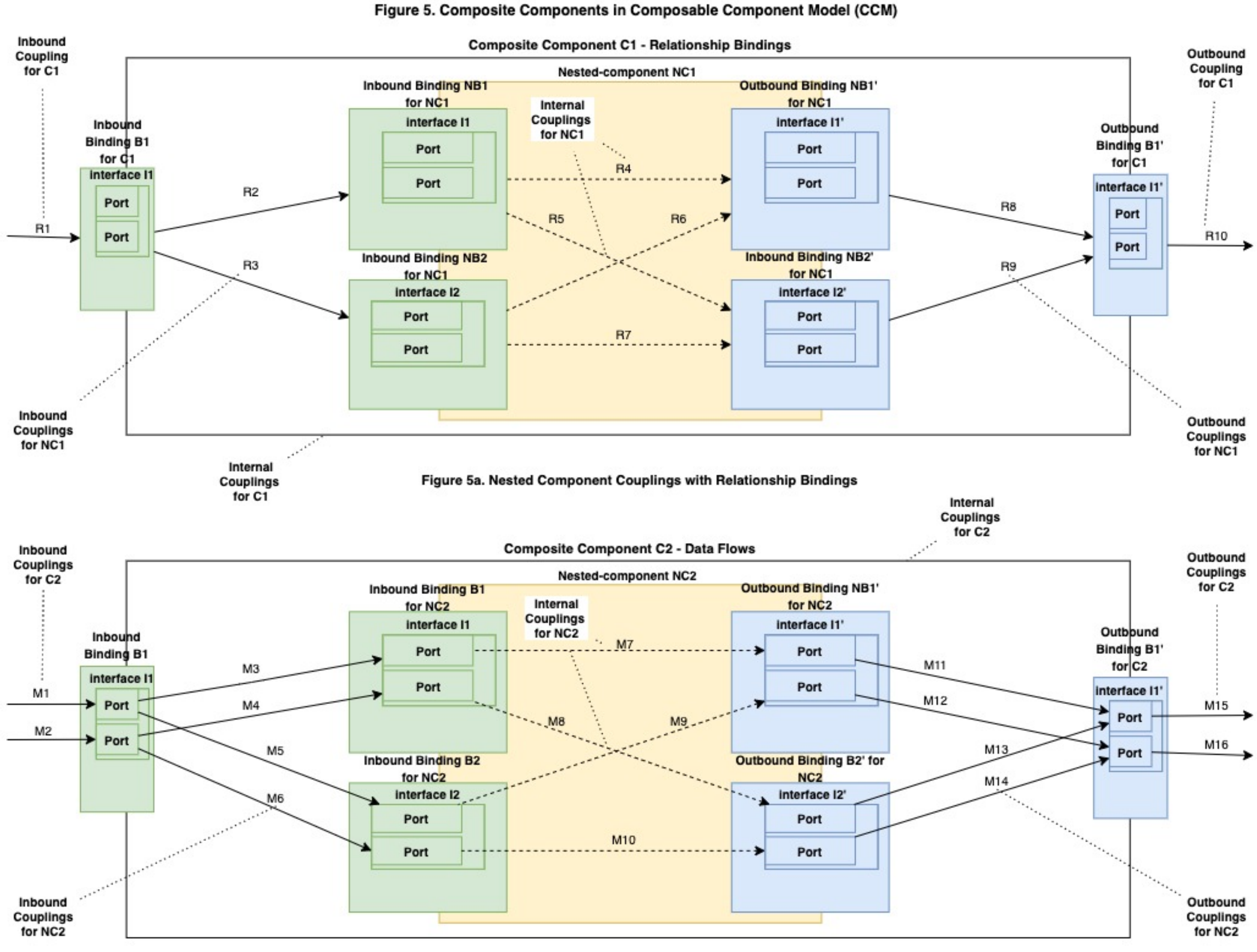

Figure 5. Composite Components in Composable Component Model (CCM)

Figure 5a. Nested Component Couplings with Relationship Bindings

Figure 5b. Nested Component Couplings with Data Flows

Modeling complex systems requires structures beyond basic components; thus, we introduce the composite component, as illustrated in Figure 5. A composite component acts as a container capable of aggregating multiple subordinate units. This container nature fundamentally changes the system's structure, as the module itself can house one or more nested components. The resulting complexity dictates that the overall system coupling is not merely additive. The total system coupling must account for three distinct interaction types: the coupling inherent to the main composite structure, the coupling that exists within the individual nested components, and the coupling that results from the interaction between these internal subunits. This necessitates a comprehensive framework to model the coupled behaviour across all hierarchical levels of the composite component.

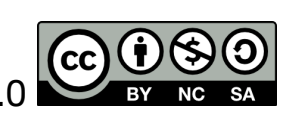

When examining composite structures like C1 (as in Figure 5a) and C2 (as in Figure 5b), the primary difference lies in the underlying interface mechanism used for inter-component communication. Both models rely on nested encapsulation—meaning the overall structure's functionality depends on the internal, coupled behaviour of its encapsulated units. However, C1 models coupling based on general structural adjacency, where the composite's overall operation is dictated by the collective interaction of its contained modules. In contrast, C2 models coupling using explicit data flow semantics. This mechanism dictates that communication is formalized through message passing; specifically, the outbound interface requires a defined message to traverse the boundary and stimulate the functionality of the connected component. Therefore, while both models maintain structural encapsulation, C1 relies on general structural coupling, whereas C2 enforces a rigorous, message-driven coupling contract, making the coupling highly dependent on the precise flow and semantics of transmitted data.

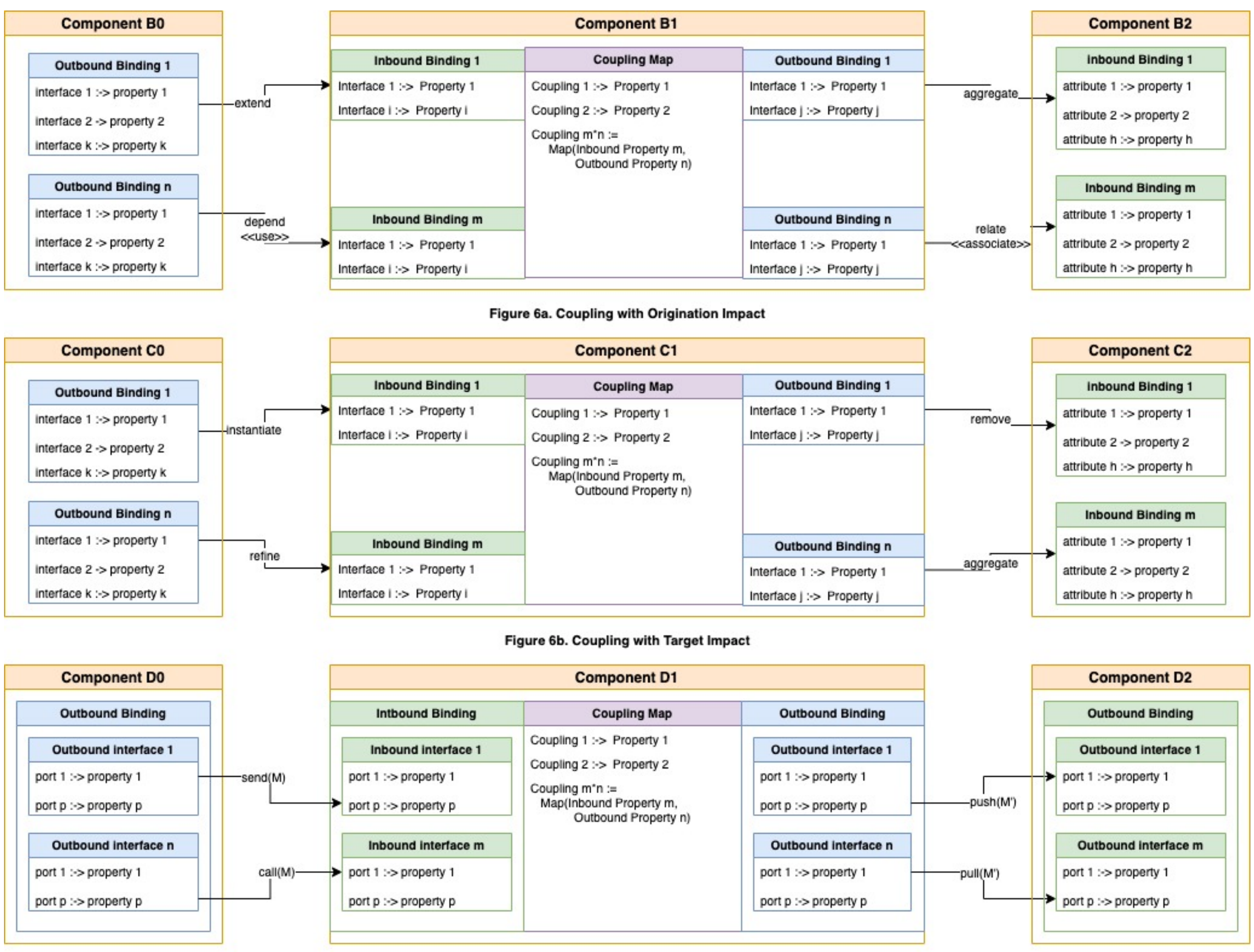

**Figure 6a. Coupling with Origination Impact**

**Figure 6b. Coupling with Target Impact**

**Figure 6c. Coupling with Bi-lateral Impact**

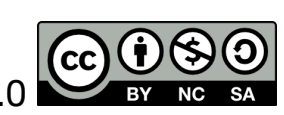

Figures 6 shows three types of coupling impacts in CCM: origination impact, target impact, and bi-lateral impact. The system's interaction is analyzed through the concept of impact, which defines the directional dependency or influence transferred between components. There are three distinct modes of impact:

Figure 6a illustrates the origination impact that describes a scenario where one component initiates a change, but only its own state is affected. The initiating component exerts influence, yet the target component remains unaffected by this action. Figure 6b illustrates the target impact that establishes a unidirectional flow where Component A directly influences Component B, and only Component B's state is affected by the external influence of A. Figure 6c illustrates the most complex state which is the bi-lateral impact, where the components exhibit mutual dependency; the modification of either component necessitates a corresponding, reciprocal change in the other.

These directional impacts—origination, target, and bi-lateral—are crucial for accurate system modeling, as they precisely delineate which component is dependent on the outcome of an interaction and how that dependency influences the overall system state.

.

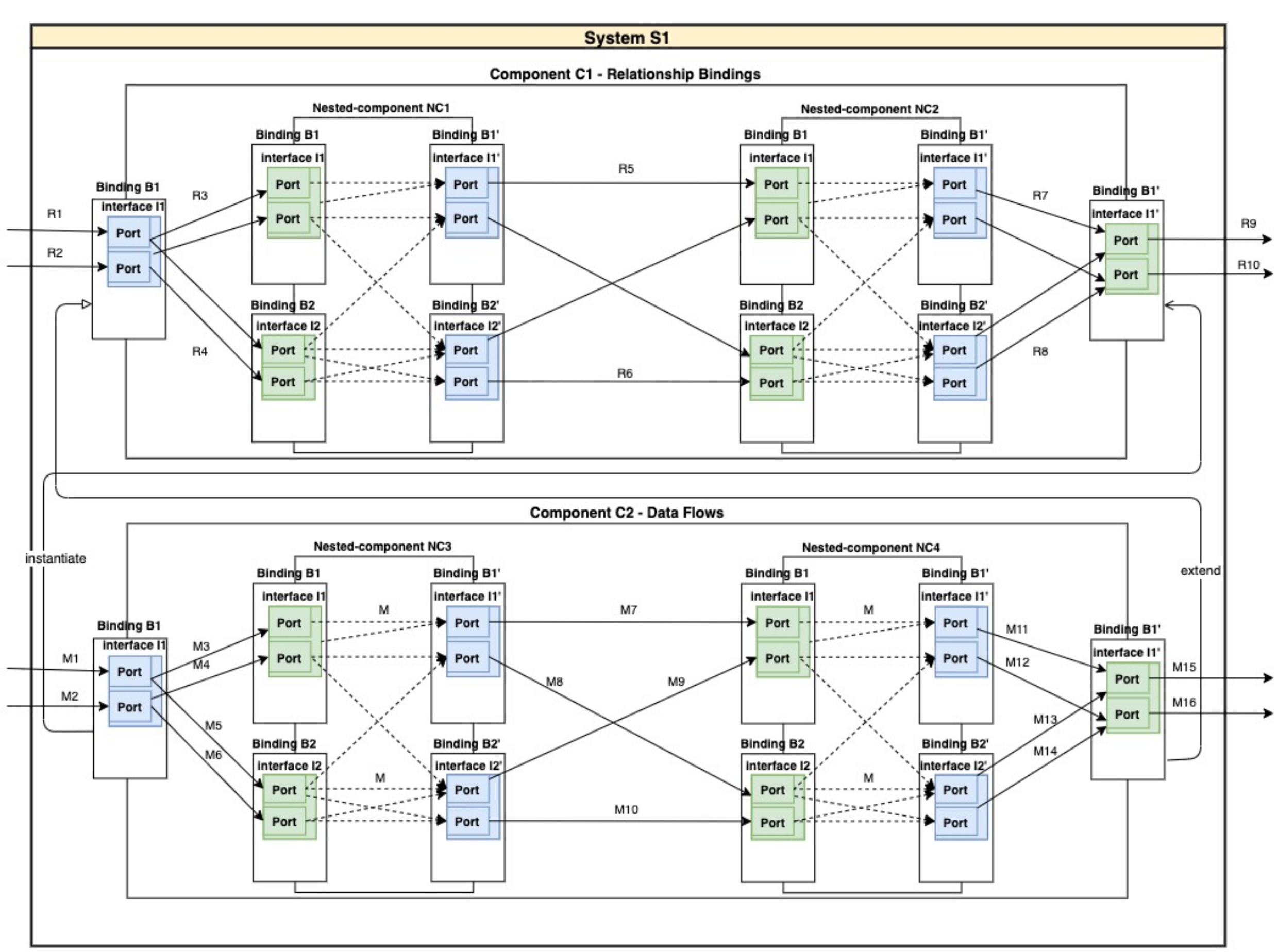

Figure 7. A Sample Composable Component System Model (CCSM)

Finally, Figure 7 illustrates the Composable Component System Model (CCSM) which is represented as a directed graph G(V,E) where nodes (V) are connected by directional edges (E). The system consists of various composable components, which are subgraphs that serve as containers for elements or other nested components.

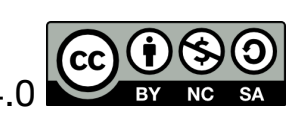

This model is highly flexible; its structural constructs allow for the representation of various topologies, including dependencies, associations, and compositions, across both instance-based and type-based relationships.

The formal foundation of our model is built upon existing complexity and coupling metrics [1, 3, 4] and system modeling standards like UML [28] and SysML [29-31]. However, the CCSM is not a general-purpose modeling language; rather, it is a specialized framework for high-level complexity and cost estimation. While the "coupling" in our model is inspired by traditional concepts, it is simplified to exclude complex control-flow and data-dependency logic, focusing instead on a streamlined architectural connection.

A key strength of the CCSM is its interoperability with existing modeling standards. Models from UML, SysML, or UCM can be mapped into the CCSM framework for analysis. By attaching numerical properties to every component, interface, and connection, the CCSM enables a quantitative approach to complexity and cost estimation, as further detailed in the next section.

## VI. Complexity Estimation for System and Components (SCCM)

While the CCSM provides a structural overview, the **System Component Complexity Model (SCCM)** provides a quantitative layer of analysis. The SCCM is built upon a formal framework of definitions and theorems (Definitions 1–12 and Theorems 1–6) used to compute the complexity of a system. In this model, every element is assigned a scalar property representing its weight or significance within a given context. By applying arithmetic operations to these scalar values, the model calculates the upper bound of system complexity.

The application of the SCCM is contingent upon the concept of **domain compatibility**. For the model to be valid, the target system must be mapped to a compatible domain where the structural elements and their relationships satisfy the underlying definitions. This allows the SCCM to be applied to various scenarios, such as the initial design of a system or the evaluation of an evolving architecture.

A critical component of this model is the analysis of **internal coupling**. In many scenarios, the internal logic of a component is opaque; therefore, the SCCM treats such components as "black boxes." In these instances, the model estimates complexity by analyzing the interaction between the component's external interfaces. Specifically, the model calculates the potential complexity by evaluating the connectivity between all inbound and outbound interfaces. By assuming a maximal connection density—where every inbound interface potentially interacts with every outbound interface—the SCCM provides a rigorous upper bound on the system's total coupling complexity, as in Figure 8.

Definition 1 (EQ1) shows that the total connectivity capacity (or the amount of couplings) of a component c, represented by its total number of interactions (CP(c)), is mathematically bounded by the sum of two distinct interaction metrics: internal coupling (INTCP) and external coupling (EXTCP).

The internal coupling, which describes the connections between elements within the component, is calculated based on the component's internal interfaces or bindings. Specifically, the maximum number of internal interactions (INTCP) is bounded by the product of the number of internal interfaces and the number of internal bindings.

Conversely, the external coupling describes all connections passing through the component's external interfaces to interact with other components. The maximum number of external interactions (EXTCP) is bounded by the sum of the number of external interfaces and the number of external bindings. These two metrics define the total connectivity capacity (couplings) of component c, governed by the relationship: CP(c) ≤ INTCP(c) + EXTCP(c).

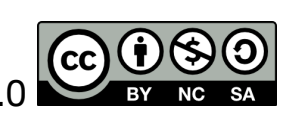

By decomposing the total connectivity into these distinct internal and external components, the model provides a precise quantification of the component's interaction potential based on the number and configuration of its interfaces and bindings.

Figure 8. System Component Complexity Model (SCCM) for Sample Basic Components

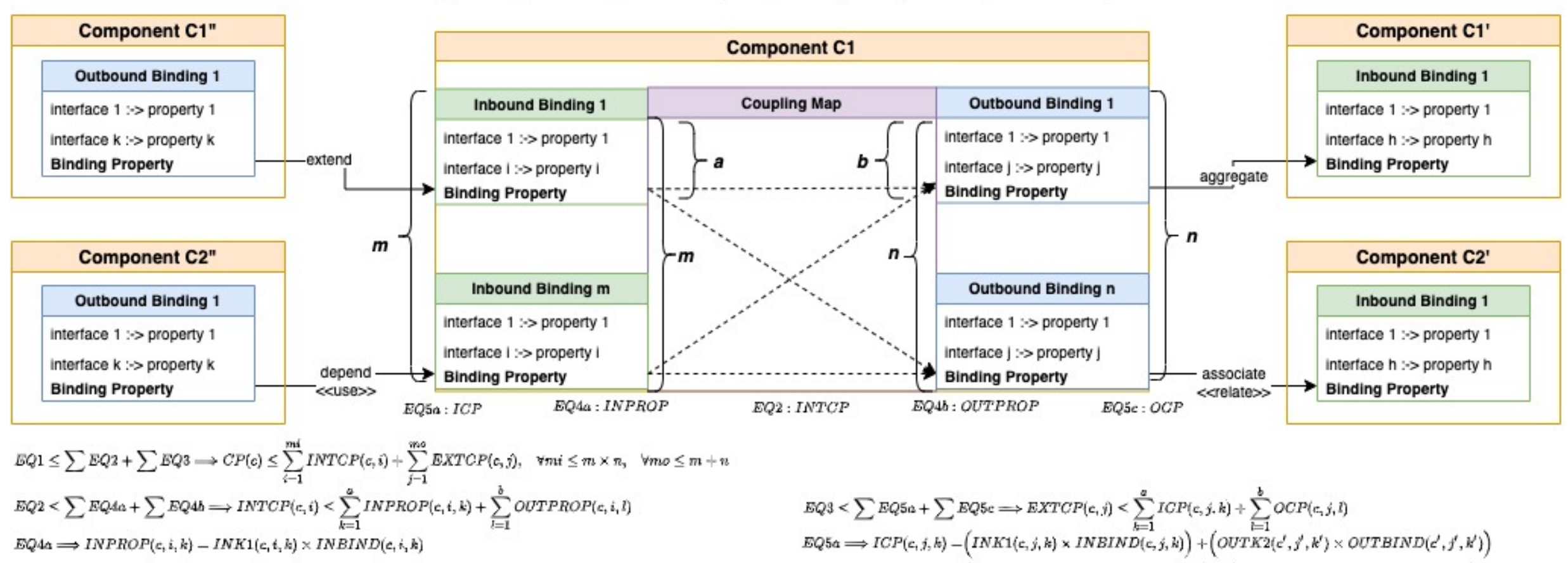

$$EQ1 \le \sum EQ2 + \sum EQ3 \Longrightarrow CP(c) \le \sum_{i=1}^{mi} INTCP(c,i) + \sum_{j=1}^{mo} EXTCP(c,j), \quad \forall mi \le m \times n, \quad \forall mo \le m + n$$

$$EQ2 < \sum EQ4a + \sum EQ4b \Longrightarrow INTCP(c,i) < \sum_{k=1}^{a} INPROP(c,i,k) + \sum_{l=1}^{b} OUTPROP(c,i,l)$$

$$EQ4a \Longrightarrow INPROP(c,i,k) - INK1(c,i,k) \times INBIND(c,i,k)$$

$$EQ4b \Longrightarrow OUTPROP(c,i,l) - OUTK1(c,i,l) \times OUTBIND(c,i,l)$$

$$EQ3 < \sum EQ5a + \sum EQ5c \Longrightarrow EXTCP(c,j) < \sum_{k=1}^{a} ICP(c,j,k) + \sum_{l=1}^{b} OCP(c,j,l)$$

$$EQ5a \Longrightarrow ICP(c,j,k) - \big(INK1(c,j,k) \times INBIND(c,j,k)\big) + \big(OUTK2(c',j',k') \times OUTBIND(c',j',k')\big)$$

$$EQ5c \Longrightarrow OCP(c,j,l) - \big(OUTK1(c,j,l) \times OUTBIND(c,j,l)\big) + \big(INK2(c',j',l') \times INBIND(c',j',l')\big)$$

Figure 8a. Internal and External Couplings with Properties of Relationship Bindings for a Basic Component

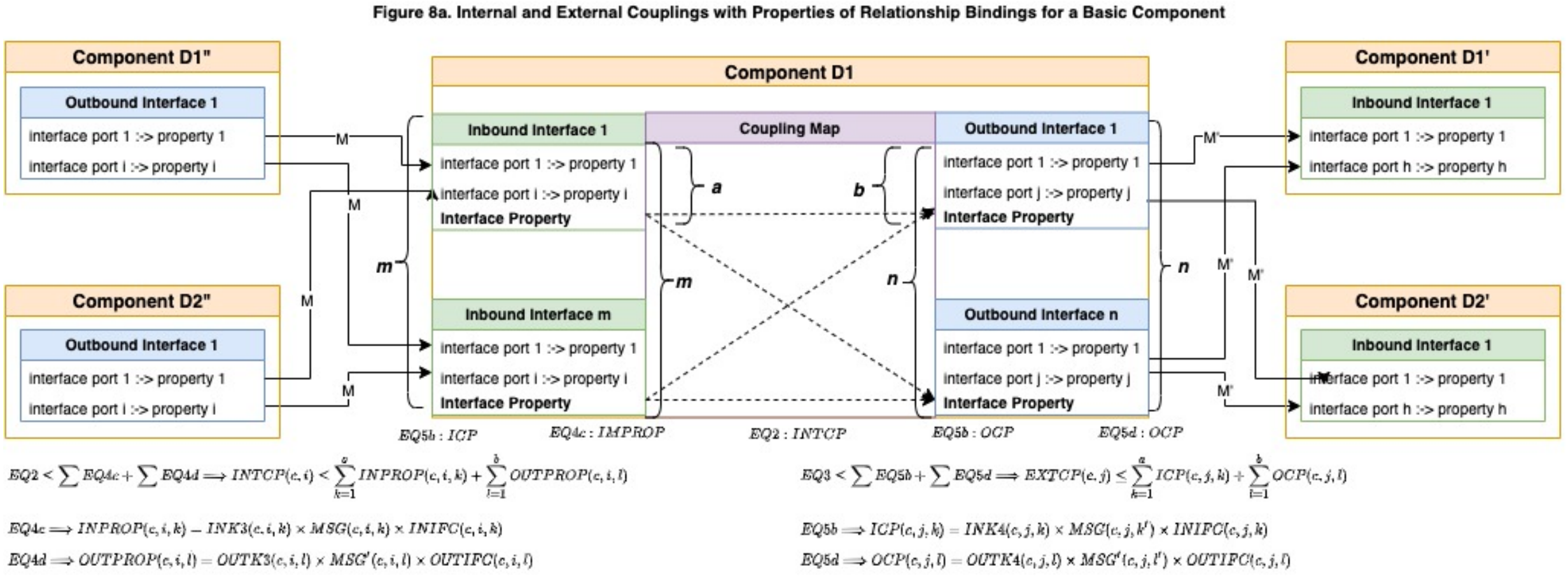

$$EQ2 < \sum EQ4c + \sum EQ4d \Longrightarrow INTCP(c,i) < \sum_{k=1}^{a} INPROP(c,i,k) + \sum_{l=1}^{b} OUTPROP(c,i,l)$$

$$EQ4c \Longrightarrow INPROP(c,i,k) - INK3(c,i,k) \times MSG(c,i,k) \times INIFC(c,i,k)$$

$$EQ4d \Longrightarrow OUTPROP(c,i,l) = OUTK3(c,i,l) \times MSG'(c,i,l) \times OUTIFC(c,i,l)$$

$$EQ3 < \sum EQ5b + \sum EQ5d \Longrightarrow EXTCP(c,j) \le \sum_{k=1}^{a} ICP(c,j,k) + \sum_{l=1}^{b} OCP(c,j,l)$$

$$EQ5b \Longrightarrow ICP(c,j,k) = INK4(c,j,k) \times MSG(c,j,k') \times INIFC(c,j,k)$$

$$EQ5d \Longrightarrow OCP(c,j,l) = OUTK4(c,j,l) \times MSG'(c,j,l') \times OUTIFC(c,j,l)$$

Figure 8b. Internal and External Couplings with Properties of Data Flows for a Basic Component

$\forall cc, cc' \in Components \quad \forall ifc = Component\ Interfaces \quad \forall imp = Component\ Implemenation$

$\forall i, i' = an\ index\ of\ Internal\ Couplings\ in\ Component\ c, c'$

$\forall j, j' = an\ index\ of\ External\ Couplings\ in\ Component\ c, c'$

$\forall k, k' = an\ index\ of\ Inbound\ Bindings\ or\ Interface\ Ports\ in\ Component\ c, c'$

$\forall l, l' = an\ index\ of\ Outbound\ Bindings\ or\ Interface\ Ports\ in\ Component\ c, c'$

$\forall m = Total\ Number\ of\ Inbound\ Bindings\ or\ Interface\ Ports\ in\ Component\ c$

$\forall n = Total\ Number\ of\ Outbound\ Bindings\ or\ Interface\ Ports\ in\ Component\ c$

$\forall a = The\ Number\ of\ Inbound\ Couplings\ connected\ to\ an\ Inbound\ Binding\ or\ Interface\ Port\ k\ in\ Component\ c$

$\forall b = The\ Number\ of\ Outbound\ Couplings\ connected\ to\ an\ Outbound\ Binding\ or\ Interface\ Port\ l\ in\ Component\ c$

$$Definition\ 1:\ CP(c) \le \sum_{i=1}^{mi} INTCP(c,i) + \sum_{j=1}^{mo} EXTCP(c,j), \quad \forall mi \le m \times n, \quad \forall mo \le m + n$$

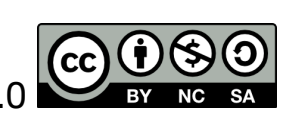

Definition 2 (EQ2) shows that the amount of an *internal coupling i (INTCP)* for a given *basic component c* is less than or equal to the sum of: the *aggregated inbound properties (INPROP)* at *inbound binding (interface port) k,* and the *aggregated outbound properties (OUTPROP)* at *outbound binding (interface port) l.* While the maximum number of external inbound couplings at an inbound binding is *a*, the maximum number of external outbound couplings at an outbound binding is *b*. An inbound binding acts as an aggregator of its external inbound couplings from other components. By *connecting every inbound coupling k to the associated outbound couplings l* within *basic component c,* the *internal coupling map acts as a mixing or fanning switch box in basic component c*.

$$Definition\ 2:\ INTCP(c,i) \leq \sum_{k=1}^{a} INPROP(c,i,k) + \sum_{l=1}^{b} OUTPROP(c,i,l)$$

Definition 3 (EQ3) shows that the amount of an *external coupling i (EXTCP)* for a given *basic component c* is less than or equal to the sum of: the amount of the *external inbound couplings (ICP)* at *inbound binding (interface port) k,* and the amount of the e*xternal outbound couplings (OCP)* at *outbound binding (interface port) l.* While the maximum number of external inbound couplings at an inbound binding is *a*, the maximum number of external outbound couplings at an outbound binding is *b*. An inbound binding (interface port) acts as an aggregator of its external inbound couplings from other components. In turn, the internal coupling map acts as a mixing or fanning switch box, which connects an external inbound coupling at *inbound binding k* with one or more external outbound couplings at *outbound binding l* within *basic component c.*

$$Definition\ 3:\ EXTCP(c,j) \leq \sum_{k=1}^{a} ICP(c,j,k) + \sum_{l=1}^{b} OCP(c,j,l)$$

The scope of the connectivity is defined by external couplings, which are further constrained by internal couplings. The functional value of these relationships is determined by three primary considerations: the pairing of external versus internal coupling, the nature of the connection (i.e., whether the relationship is mediated by a message or a direct connection), and the calculated impact of the association. The quantification of this impact follows specific formulae: the initial assessment sums the weighted values of the linked elements; alternatively, if the connection is mediated by a message, the impact is calculated by summing the weighted value of the message and the associated element. Furthermore, the absolute value assigned to any element is generally proportional to its complexity, and for practical modeling purposes, this complexity is frequently determined by counting the number of associated attributes.

For relational couplings, Definition 4a (EQ4a) shows that an *aggregated inbound properties k (INPROP) of an internal coupling i* in *basic component c* is equal to the product of: the weighted constant (*INK1)* and the property value of an *inbound binding (INBIND).* Definition 4b (EQ4b) shows that an *aggregated outbound properties l (OUTPROP)) of an internal coupling i* in *basic component c* is equal to the product of: the weighted constant *(OUTK1)* and the property value of an *outbound binding (OUTBIND).*

For data-flow couplings, Definition 4c (EQ4c) shows that an *aggregated inbound properties k (INPROP) of an internal coupling i* in *basic component c* is equal to the product of: the weighted constant (*INK3)*, the property value of an *inbound message (MSG),* and the property value at *port k* of *inbound interface i (INIFC)* of *basic component c*. Definition 4d (EQ4d) shows that an *aggregated outbound properties l (OUTPROP)) of an internal coupling i* in *basic component c* is equal to the product of: the weighted constant *(OUTK3),* the property value of an *outbound message (MSG'),* and the property value at *port l* of *outbound interface i (OUTIFC)* of *basic component c.*

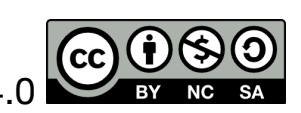

$$Definition\ 4a:\ INPROP(c,i,k) = INK1(c,i,k) \times INBIND(c,i,k)$$
$$Definition\ 4b:\ OUTPROP(c,i,l) = OUTK1(c,i,l) \times OUTBIND(c,i,l)$$
$$Definition\ 4c:\ INPROP(c,i,k) = INK3(c,i,k) \times MSG(c,i,k) \times INIFC(c,i,k)$$
$$Definition\ 4d:\ OUTPROP(c,i,l) = OUTK3(c,i,l) \times MSG'(c,i,l) \times OUTIFC(c,i,l)$$

For relational couplings, Definition 5a (EQ5a) shows that an *external inbound coupling k (ICP) of an external coupling j* in *basic component c* is equal to the sum of: the product of the weighted constant (OUT*K1)* and the property value of an *inbound binding (INBIND),* and the product of the weighted constant *(OUTK2)* and the property value of an o*utbound binding (OUTBIND).* Definition 5c (EQ5c) shows that an *external outbound coupling l (OCP) of an external coupling j* in *basic component c* is equal to the sum of: the product of the weighted constant *(OUTK1)* and the property value of an out*bound binding (OUTBIND)* of *basic component c,* and the product of the weighted constant *(INK2)* and the property value of an *inbound binding (INBIND) of external component c'.*

For data-flow couplings, Definition 5b (EQ5b) shows that an *external inbound coupling l (ICP) of an external coupling j* in *basic component c* is equal to the product of: the weighted constant (OUT*K4),* the *property value of an outbound message (MSG),* and the property value at *port k* of *inbound interface i (INIFC)* of *basic component c.* Definition 5d (EQ5d) shows that an *external outbound coupling l (OCP) of an external coupling j* is equal to the product of: the weighted constant *(OUTK4),* the property value of an *outbound message (MSG'),* and the property value at *port l* of *outbound interface i (OUTIFC)* of *basic component c.*

$$Definition\ 5a:\ ICP(c,j,k) = \left(INK1(c,j,k) \times INBIND(c,j,k)\right) + \left(OUTK2(c',j',k') \times OUTBIND(c',j',k')\right)$$
$$Definition\ 5b:\ ICP(c,j,k) = INK4(c,j,k) \times MSG(c,j,k') \times INIFC(c,j,k)$$
$$Definition\ 5c:\ OCP(c,j,l) = \left(OUTK1(c,j,l) \times OUTBIND(c,j,l)\right) + \left(INK2(c',j',l') \times INBIND(c',j',l')\right)$$
$$Definition\ 5d:\ OCP(c,j,l) = OUTK4(c,j,l) \times MSG'(c,j,l') \times OUTIFC(c,j,l)$$

A composite component is a specialized type of component that integrates multiple nested components within its structure. This integration capability fundamentally expands the component's overall functionality and connectivity. To quantify this complexity, Definition 6a shows that the total number of external inbound couplings of c*omposite component cc* is less than or equal to the product of: the number of its inbound bindings, the number of the outbound bindings of its external components, and the number of its external components.

Definition 6b shows that the total number of external outbound couplings of c*omposite component cc* is less than or equal to the product of: the number of its outbound bindings, the number of the inbound bindings of its external components, and the number of its external components.

Definition 6c shows that the total number of external couplings of c*omposite component cc* with its external components is less than or equal to the sum of the numbers of its external inbound and outbound bindings.

Definition 6d shows that the total number of internal couplings of the nested components in c*omposite component cc* is less than or equal to the product of the number of its inbound bindings, the number of outbound bindings, and the number of its nested components.

Definition 6e shows that the total number of external couplings between the nested components in c*omposite component cc* is less than or equal to the product of: the sum of its inbound and outbound bindings and the number of its nested components.

Definition 6f shows that the total number of external couplings from the nested components to the c*omposite component cc* is less than or equal to the product of: the sum of its inbound and outbound bindings and the square of the number of its interconnecting nested components.

$\forall cc, cc' \in Composite\ Components; \quad \forall ifc = Component\ Interfaces \quad \forall imp = Component\ Implemenation$

$\forall exicp = The\ Number\ of\ External\ Inbound\ Couplings\ of\ Composite\ Component\ cc\ with\ its\ External\ Components$

$\forall exocp = The\ Number\ of\ External\ Outbound\ Couplings\ of\ Composite\ Component\ cc\ with\ its\ External\ Components$

$\forall excp = The\ Number\ of\ External\ Couplings\ of\ Composite\ Component\ cc\ with\ its\ External\ Components$

$\forall incp = The\ Number\ of\ Internal\ Couplings\ of\ Nested\ Components\ in\ Composite\ Component\ cc$

$\forall necp = Total\ Number\ of\ External\ Couplings\ between\ Nested\ Components\ in\ Composite\ Component\ cc$

$\forall nccp = Total\ Number\ of\ External\ Couplings\ from\ Nested\ Components\ to\ Composite\ Component\ cc$

$\forall ex = Total\ Number\ of\ External\ Components\ for\ Composite\ Component\ cc$

$\forall nc = Total\ Number\ of\ Nested\ Components\ for\ Composite\ Component\ cc$

$\forall ci\ =\ an\ index\ of\ Nested\ Components\ in\ Component\ cc$

$\forall ni\ =\ an\ index\ of\ Internal\ Couplings\ of\ Nested\ Components\ in\ Component\ cc$

$\forall nj\ =\ an\ index\ of\ External\ Couplings\ from\ Nested\ Components\ to\ Component\ cc$

$\forall nk\ =\ an\ index\ of\ External\ Couplings\ between\ Nested\ Components\ in\ Component\ cc$

$\forall nl\ =\ an\ index\ of\ External\ Couplings\ from\ Component\ cc\ to\ its\ External\ Components$

$\forall m, m' = Total\ Number\ of\ Inbound\ Bindings\ or\ Interface\ Ports\ for\ any\ Components\ or\ Systems$

$\forall n, n' = Total\ Number\ of\ Outbound\ Bindings\ or\ Interface\ Ports\ for\ any\ Components\ or\ Systems$

$Definition\ 6a:\ exicp \le (m \times n' \times ex)$

$Definition\ 6b:\ exocp \le (n \times m' \times ex)$

$Definition\ 6c:\ excp \le exicp + exocp$

$Definition\ 6d:\ incp \le (m \times n) \times nc$

$Definition\ 6e:\ necp \le (m + n) \times nc$

$Definition\ 6f:\ nccp \le (m + n) \times nc^2$

We derive Definition 7a using Definitions 1-6 to show that the total amount of couplings of a *composite component cc (CP(cc))* is less than or equal to the sum of: the amount of the internal couplings of its nested components *(INTCP(cc, ni'))*, the amount of the external couplings between its nested components *(EXTCP(cc, nj))*, the amount of the external couplings from its nested components *(INTCP(cc, nk))* to the composite component cc, and the amount of the external couplings of composite component cc with its external components (*EXTCP(cc, nl))*.

Definitions 7b shows that the total amount of couplings of the nested *component cc (CP(ci))* in composite component cc is less than or equal to the sum of: the amount of the internal couplings of its nested components *(INTCP(cc, ni'))*, the amount of the external couplings between its nested components *(EXTCP(cc, nj))*, and the amount of the external couplings from its nested components *(INTCP(cc, nk))* to the composite component cc.

By combining Definitions 7a and 7b, we arrive to *Theorem 1* that shows the total amount of couplings of a composite component cc is less than or equal to the sum of the couplings of its nested components and the external couplings with its external components.

$$Definition\ 7a:\ CP(cc) \le \left( \sum_{ni=1}^{incp} INTCP(cc, ni) + \left( \sum_{nj=1}^{necp} EXTCP(cc, nj) + \sum_{nk=1}^{nccp} EXTCP(cc, nk) \right) \right) + \sum_{nl=1}^{excp} EXTCP(cc, nl)$$

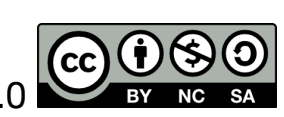

$$Definition\ 7b: \sum_{ci=1}^{nc} CP(ci) \leq \left( \sum_{ni=1}^{incp} INTCP(cc, ni) + \left( \sum_{nj=1}^{necp} EXTCP(cc, nj) + \sum_{nk=1}^{nccp} EXTCP(cc, nk) \right) \right)$$

$$\textbf{Theorem 1}: CP(cc) \leq \sum_{ci=1}^{nc} CP(ci) + \sum_{nl=1}^{excp} EXTCP(cc, nl)$$

Definitions 8-9 describe the magnitude of complexity as couplings of the APIs and implementation of a component. Definition 8a indicates that the complexity of the *APIs (ifc)* of a *component c* is less than or equal to the sum of its external inbound couplings. Definition 8b demonstrates that the magnitude of complexity of the *implementation (imp)* of a *component c* is less than or equal to the sum of its internal couplings and the sum of its external outbound couplings. Definition 9a shows that the the magnitude of complexity of the *APIs (ifc)* of a *composite component cc* is less than or equal to the sum of its external inbound couplings. Definition 9b shows that the magnitude of complexity of the *implementation (imp)* of a *composite component cc* is less than or equal to the sum of the couplings of its nested components and its external outbound couplings with its external components.

$$Definition\ 8a: CP(c, ifc) \leq \sum_{j=1}^{m} EXTCP(c, j)$$

$$Definition\ 8b: CP(c, imp) \leq \sum_{i=1}^{mi} INTCP(c, i) + \sum_{j=1}^{n} EXTCP(c, j), \quad \forall mi \leq m \times n$$

$$Definition\ 9a: CP(cc, ifc) \leq \sum_{nl=1}^{exicp} EXTCP(cc, nl)$$

$$Definition\ 9b: CP(cc, imp) \leq \sum_{ci=1}^{nc} CP(ci) + \sum_{nl=1}^{exocp} EXTCP(cc, nl)$$

Given that a component consisting of component interfaces and implementation, we arrive to *Theorem 2* based on Definitions 8a-8b to show that the complexity of a *component c (CP(c))* is equal to the sum of its *interface complexity (CP(c, ifc))* and its *implementation complexity (CP(c, imp))*. Similarly, we arrive to *Theorem 3* based on Definitions 9a-9b to show that the complexity of a *composite component cc (CP(cc))* is equal to the sum of its *interface complexity (CP(cc, ifc))* and its *implementation complexity (CP(cc, imp))*.

$$\textbf{Theorem 2}: CP(c) = CP(c, ifc) + CP(c, imp)$$

$$\textbf{Theorem 3}: CP(cc) = CP(cc, ifc) + CP(cc, imp)$$

Similar to a composite component, a self-contained system [27] consists of system-level components (system components), which are nested components within a system, as in Figure 7. A system component can be either a basic or composite component. Similar to a nested component within a composite component, a system component has its internal couplings, its external couplings with the system, and its external couplings with other system components. Based on Definitions 7a, 7b, and 10a-f, we arrive to *Theorem 4* to show that the total amount of couplings of a *System sys (CP(sys))* is less than or equal to the sum of: the internal couplings of its system components *(INTCP(sys, si))*, the the external couplings between its system components *(EXTCP(sys, sj))*, and the external couplings from its system components *(INTCP(sys, sk))* to the system, and its external couplings to its external systems *(EXTCP(sys, sl))*. Next, we derive Definition 11 based on Definition 7b and Theorem 4. Then, we use Definition 11 to arrive to *Theorem 5,* which states that the complexity of a *system (CP(sys))* is less than or equal to the sum of the couplings of its system components and its external couplings with other external systems.

[1]This material can only be used for non-commercial purposes. Please contact the author for commercial licensing terms.

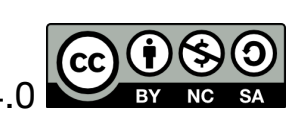

$\forall sys = System \quad \forall ifc = System\ Interfaces \quad \forall imp = System\ Implemenation$
$\forall sxicp = The\ Number\ of\ External\ Inbound\ Couplings\ of\ System\ sys\ with\ its\ External\ Systems$
$\forall sxocp = The\ Number\ of\ External\ Outbound\ Couplings\ of\ System\ sys\ with\ its\ External\ Systems$
$\forall sxcp = The\ Number\ of\ External\ Couplings\ for\ System\ sys\ with\ its\ External\ Systems$
$\forall sincp = The\ Number\ of\ Internal\ Couplings\ of\ System\ Components\ in\ System\ sys$
$\forall secp = Total\ Number\ of\ External\ Couplings\ between\ System\ Components\ in\ System\ sys$
$\forall sccp = Total\ Number\ of\ External\ Couplings\ from\ System\ Components\ to\ System\ sys$
$\forall sci = an\ index\ of\ System\ Components\ in\ System\ sys$
$\forall se = Total\ Number\ of\ External\ Systems\ for\ System\ sys$
$\forall sc = Total\ Number\ of\ System\ Components\ in\ System\ sys$
$\forall si = an\ index\ of\ Internal\ Couplings\ of\ System\ Components\ in\ System\ sys$
$\forall sj = an\ index\ of\ External\ Couplings\ from\ System\ Components\ to\ System\ sys$
$\forall sk = an\ index\ of\ External\ Couplings\ between\ System\ Components\ in\ System\ sys$
$\forall sl = an\ index\ of\ External\ Couplings\ from\ System\ sys\ to\ its\ External\ Systems$

$Definition\ 10a:\ sxicp \leq (m \times n' \times se)$
$Definition\ 10b:\ sxocp \leq (n \times m' \times se)$
$Definition\ 10c:\ sxcp \leq (m \times n' \times se) + (n \times m' \times se)$
$Definition\ 10d:\ sincp \leq (m \times n) \times sc$
$Definition\ 10e:\ secp \leq (m + n) \times sc^2$
$Definition\ 10f:\ sccp \leq (m + n) \times sc$

$$\textbf{Theorem 4}:\ CP(sys) \leq \left(\sum_{si=1}^{sincp} INTCP(sys, si) + \left(\sum_{sj=1}^{secp} EXTCP(sys, sj) + \sum_{sk=1}^{sccp} EXTCP(sys, sk)\right)\right) + \sum_{sl=1}^{sxcp} EXTCP(sys, sl)$$

$$Definition\ 11:\ \sum_{sci=1}^{sc} CP(sci) \leq \left(\sum_{si=1}^{sincp} INTCP(sys, si) + \left(\sum_{sj=1}^{secp} EXTCP(sys, sj) + \sum_{sk=1}^{sccp} EXTCP(sys, sk)\right)\right)$$

$$\textbf{Theorem 5}:\ CP(sys) \leq \sum_{sci=1}^{sc} CP(sci) + \sum_{sl=1}^{sxcp} EXTCP(sys, sl)$$

We formulate Definition 12a based on Definition 9a to show that the magnitude of complexity of the *APIs (ifc)* of a *system sys* is less than or equal to the sum of its inbound external couplings. We also Definition 12b based on Definition 9b to show that the magnitude of complexity of the *implementation (imp)* of a *system sys* is less than or equal to the sum of the couplings of its system components and the external outbound couplings with other external systems. Given that a system consisting of system interfaces and implementation, we derive *Theorem 6* from Definitions 12a, 12b, and Theorem 3 to show that the magnitude of complexity of a *system sys* is less than or equal to the sum of its *interface complexity (CP(sys, ifc))* and its *implementation complexity (CP(sys, imp))*.

$$Definition\ 12a:\ CP(sys, ifc) \leq \sum_{sl=1}^{sxicp} EXTCP(sys, sl)$$

$$Definition\ 12b:\ CP(sys, imp) \leq \sum_{sci=1}^{sc} CP(sci) + \sum_{sl=1}^{sxocp} EXTCP(sys, sl)$$

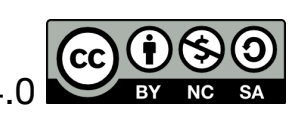

**Theorem 6** : $CP(sys) = CP(sys, ifc) + CP(sys, imp)$

Our coupling and complexity model is informed by existing research, drawing inspiration from methodologies such as Cardoso's Control Flow Complexity [3], Quynh and Thang's Dynamic Coupling Metrics [1], and Yang's Indirect Coupling [4]. Critically, this model improves upon prior work by avoiding the need for deep inspection of internal component mechanics. Because it operates without requiring granular, low-level analysis of internal logic, it remains highly flexible and robust across varied systems..

## VII. Cost Estimation for System and Components (SCEM)

Building upon the foundation of the SCCM, we introduce the System Cost Estimation Model (SCEM). The SCEM serves as an extension of the SCCM, augmenting the existing complexity framework with a cost-centric dimension. While the SCCM characterizes the structural complexity of a system, the SCEM incorporates the economic variables—such as labor effort and monetary expenditure—required for its realization. For instance, whereas an SCCM analysis might evaluate the architecture of a payment system, the SCEM extends this to estimate the human and financial resources necessary for its development. This progression is formally realized by expanding our mathematical primitives, moving from Definitions 1–12 and Theorems 1–6 to the more comprehensive set of Definitions 13–25 and Theorems 7–15.

Definition 13 shows that the total cost of a component c (*CTS)* is less than or equal to the total cost of its elements and the total cost of its couplings *(CPCTS)*. Definition 14 shows that the cost of an element *(ELMCTS)* is equal to its cost factor *(CFELM)* multiplied by the property value of the element *(ELEMENT)*. Definition 15 shows that the cost of the couplings of a component c *(CPCTS)* is less than or equal to the sum of the costs of its internal couplings *(INTCTS)* and external couplings *(EXTCTS)*. Definition 16 shows that the cost of an internal coupling *(INTCTS)* is less than or equal to the sum of the costs of its *aggregated inbound properties (INPCTS)* and *aggregated outbound properties (OUTPCTS)*. Definition 17 shows that the cost of an external coupling *(EXTCTS)* is less than or equal to the sum of the costs of its external inbound couplings *(ICTS)* and external outbound couplings *(OCTS)*.

$$Definition\ 13:\ CTS(c) \le \sum_{j=1}^{ce} ELMCTS(j) + CPCTS(c),\quad \forall ce = m + n$$

$$Definition\ 14:\ ELMCTS(j) = CFELM(j) \times ELEMENT(j)$$

$$Definition\ 15:\ CPCTS(c) \le \sum_{i=1}^{mi} INTCTS(c,i) + \sum_{j=1}^{mo} EXTCTS(c,j),\quad \forall mo \le m + n$$

$$Definition\ 16:\ INTCTS(c,i) \le \sum_{k=1}^{a} INPCTS(c,i,k) + \sum_{l=1}^{b} OUTPCTS(c,i,l)$$

$$Definition\ 17:\ EXTCTS(c,j) \le \sum_{k=1}^{a} ICTS(c,j,k) + \sum_{l=1}^{b} OCTS(c,j,l)$$

Definition 18a shows that the cost of an *aggregated inbound property (INPCTS)* is equal to the product of its cost factor *(CFPICP)* and the amount of its aggregated inbound property *(INPROP)*. Definition 18b shows that the cost of an external inbound coupling is equal to the product of its cost factor *(CFICP)* and the amount of the external inbound coupling *(ICP)*. Definition 19a shows that the cost of an *aggregated outbound property (OUTPCTS)* is equal to the product of its cost factor *(CFOUTP)* and the amount of its *aggregated outbound property (OUTPROP)*. Definition 19b shows that the cost of an external outbound coupling for external couplings is equal to the product of its cost factor *(CFOCP)* and the amount of the external outbound coupling *(OCP)*.

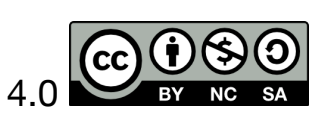

$$Definition\ 18a:\ INPCTS(c,i,k) = CFINP(c,i,k) \times INPROP(c,i,k)$$
$$Definition\ 18b:\ ICTS(c,i,k) = CFICP(c,i,k) \times ICP(c,i,k)$$
$$Definition\ 19a:\ OUTPCTS(c,j,) = CFOUTP(c,j,l) \times OUTPROP(c,j,l)$$
$$Definition\ 19b:\ OCTS(c,j,l) = CFOCP(c,j,l) \times OCP(c,j,l)$$

Definition 20a shows that the cost of a composite component *(CPCTS(cc))* is less than or equal to the sum of: the total cost of the internal couplings of its nested components *(INTCTS(cc,ni))*, the total cost of the external couplings between its nested components *(EXTCTS(cc,nj)),* the total cost of the external couplings from its nested components to the composite component cc *(EXTCTS(cc,nk))*, and the total cost of its external couplings *(EXTCTS(cc,nl))*.

Definition 20b shows that the total cost of the couplings of the nested components *(CPCTS(ci))* in *composite component cc* is less than or equal to the sum of: the total cost of the internal couplings of its nested components *(INTCTS(cc,ni))*, the total cost of the external couplings between its nested components *(EXTCTS(cc,nj)),* and the total cost of the external couplings from its nested components to the composite component cc *(EXTCTS(cc,nk))*. Therefore, we arrive to *Theorem 8* that shows the total cost of couplings of a *composite component cc* is less than equal to the sum of the total cost of the couplings of the nested components *(CPCTS(ci))* and the total cost of its external couplings *(EXTCTS(cc,nl))*. *Theorem 9* shows that the cost of the *composite component cc* is less than or equal to the total cost of its nested components *(CPCTS(ci))* and the total cost of its couplings *(CPCTS(cc))*.

$$Definition\ 20a:\ CPCTS(cc) \le \left( \sum_{ni=1}^{incp} INTCTS(cc,ni) + \left( \sum_{nj=1}^{necp} EXTCTS(cc,nj) + \sum_{nk=1}^{nccp} EXTCTS(cc,nk) \right) \right) + \sum_{nl=1}^{excp} EXTCTS(cc,nl)$$

$$Definition\ 20b:\ \sum_{ci=1}^{nc} CPCTS(ci) \le \left( \sum_{ni=1}^{incp} INTCTS(cc,ni) + \left( \sum_{nj=1}^{necp} EXTCTS(cc,nj) + \sum_{nk=1}^{nccp} EXTCTS(cc,nk) \right) \right)$$

$$\textbf{Theorem 8}:\ CPCTS(cc) \le \sum_{ci=1}^{nc} CPCTS(ci) + \sum_{nl=1}^{excp} EXTCTS(cc,nl)$$

$$\textbf{Theorem 9}:\ CTS(cc) \le \sum_{ci=1}^{nc} CTS(ci) + CPCTS(cc)$$

Definitions 22a-23b describe the costs of the APIs and implementation of a component. They are built upon *Definition 21,* Theorems 8 and 9, and Definitions 20 and 20b. *Definition 22a* shows that the cost of the *APIs (ifc)* of *component c (CTS(c,ifc))* is less than or equal to the sum of the costs of the external inbound couplings *(EXTCTS)* and the costs of the inbound elements *(ELMCTS)*. *Definition 22b* shows that the cost of the *implementation (imp)* of *component c (CTS(c,imp))* is less than or equal to the sum of: the costs of the internal couplings *(INTCTS)* of its nested components, the costs of the external outbound couplings *(EXTCTS)*, and the costs of its outbound elements *(ELMCTS)*. *Definition 23a* indicates that the cost of the *APIs (ifc)* of *composite component cc (CTS(cc,ifc))* is less than or equal to the sum of the costs of the external inbound couplings *(EXTCTS)* and the costs of its inbound elements *(ELMCTS)*. *Definition 23b* shows that the cost of the *implementation (imp)* of *composite component cc*

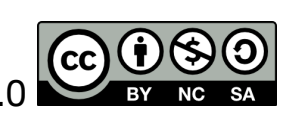

*(CTS(cc,imp))* is less than or equal to the sum of: the costs of its nested components *(CTS(ci))*, the costs of the external outbound couplings *(EXTCTS)*, and the costs of its outbound elements *(ELMCTS)*.

$$Definition\ 21:\ CTS(cc) \le \left(\sum_{ci=1}^{nc} CPCTS(ci) + \sum_{nl=1}^{excp} EXTCTS(cc,nl)\right) + CPCTS(cc)$$

$$Definition\ 22a:\ CTS(c,ifc) \le \sum_{j=1}^{m} EXTCTS(c,j) + \sum_{j=1}^{m} ELMCTS(j)$$

$$Definition\ 22b:\ CTS(c,imp) \le \sum_{i=1}^{mi} INTCTS(c,i) + \sum_{j=1}^{n} EXTCTS(c,j) + \sum_{j=1}^{n} ELMCTS(j),\ \ \forall mi \le m \times n$$

$$Definition\ 23a:\ CTS(cc,ifc) \le \sum_{nl=1}^{exicp} EXTCTS(cc,nl) + \sum_{j=1}^{m} ELMCTS(j)$$

$$Definition\ 23b:\ CTS(cc,imp) \le \sum_{ci=1}^{nc} CTS(ci) + \sum_{nl=1}^{exocp} EXTCTS(cc,nl) + \sum_{j=1}^{n} ELMCTS(j),\ \ \forall mi \le m \times n$$

*Theorem 10* indicates that the cost of a *component c (CTS(c))* is less than or equal to the sum of the cost of its interface complexity (CTS(c,ifc)) and the costs of its implementation complexity *(CTS(c,imp))*. *Theorem 11* indicates that the cost of a *composite component cc (CTS(cc))* is less than or equal to the sum of the cost of its interface complexity *(CTS(cc,ifc))* and the costs of its implementation complexity *(CTS(cc,imp))*.

**Theorem 10**: $CTS(c) = CTS(c,ifc) + CTS(c,imp)$

**Theorem 11**: $CTS(cc) = CTS(cc,ifc) + CTS(cc,imp)$

*Theorem 12* shows that the cost of the couplings of *system sys* is less than or equal to the sum of: the total cost of the internal couplings of its system components *(INTCTS(sys,si))*, the total cost of the external couplings between its system components *(EXTCTS(sys,sj)),* the total cost of the external couplings from its system components to the system sys *(EXTCTS(sys,sk))*, and the total cost of its external couplings *(EXTCTS(sys,sl))* with other systems.

Definition 24 shows that the total cost of the couplings of the system components *(CPCTS(sci))* in *system sys* is less than or equal to the sum of: the total cost of the internal couplings of its system components *(INTCTS(sys,si))*, the total cost of the external couplings between its system components *(EXTCTS(sys,sj)),* the total cost of the external couplings from its system components to the system sys *(EXTCTS(sys,sk))*. Therefore, we arrive to *Theorem 13* that shows the total cost of couplings of *system sys* is less than equal to the sum of the total cost of the couplings of the system components *(CPCTS(sci))* and the total cost of its external couplings *(EXTCTS(sys,sl))* with other systems. *Theorem 14* shows that the cost of *system sys* is less than or equal to the total cost of its system components *(CPCTS(sci))* and the total cost of its couplings *(CPCTS(sys))*.

**Theorem 12**: $CPCTS(sys) \le \left(\sum_{si=1}^{sincp} INTCTS(sys,si) + \left(\sum_{sj=1}^{secp} EXTCTS(sys,sj) + \sum_{sk=1}^{sccp} EXTCTS(sys,sk)\right)\right) + \sum_{sl=1}^{sxicp} EXTCTS(sys,sl)$

$$Definition\ 24:\ \sum_{sci=1}^{sc} CPCTS(sci) \le \left(\sum_{si=1}^{sincp} INTCTS(sys,si) + \left(\sum_{sj=1}^{secp} EXTCTS(sys,sj) + \sum_{sk=1}^{sccp} EXTCTS(sys,sk)\right)\right)$$

**Theorem 13**: $CPCTS(sys) \le \sum_{sci=1}^{sc} CPCTS(sci) + \sum_{sl=1}^{sxicp} EXTCTS(sys,sl)$

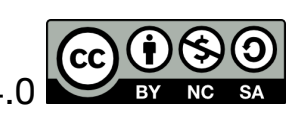

$$\textbf{Theorem 14}: CTS(sys) \leq \sum_{sci=1}^{sc} CTS(sci) + CPCTS(sys)$$

*Definition 25a* indicates that the cost *(CTS(sys,ifc))* of the *APIs (ifc)* of *system sys* is less than or equal to the sum of the costs of the external inbound couplings *(EXTCTS)* and the costs of its inbound system elements *(ELMCTS)*.

*Definition 25b* shows that the cost *(CTS(sys,imp))* of the *implementation (imp)* of *system sys* is less than or equal to the sum of: the costs of its system components *(CTS(sci))*, the costs of the external outbound couplings *(EXTCTS(sys,sl))*, and the costs of its outbound system elements *(ELMCTS)*.

$$Definition\ 25a: CTS(sys, ifc) \leq \sum_{sl=1}^{sxicp} EXTCTS(sys, sl) + \sum_{j=1}^{m} ELMCTS(j)$$

$$Definition\ 25b: CTS(sys, imp) \leq \sum_{sci=1}^{sc} CTS(sci) + \sum_{sl=1}^{sxocp} EXTCTS(sys, sl) + \sum_{j=1}^{n} ELMCTS(j)$$

Given that a system consisting of system interfaces and implementation, we derive *Theorem 15* from Definitions 25a and 25b to show that the cost of *system sys (CTS(sys))* is less than or equal to the sum of the cost of its interface complexity *(CTS(sys,ifc))* and the costs of its implementation complexity *(CTS(sys,imp))*.

$$\textbf{Theorem 15}: CTS(sys) = CTS(sys, ifc) + CTS(sys, imp)$$

The SCCM and its underlying definitions are used to construct a tabular version of the model; we apply this representation to a practical example in the next section.

## VIII. Apply Graphical and Tabular Form To a Sample Retail Banking System Integration

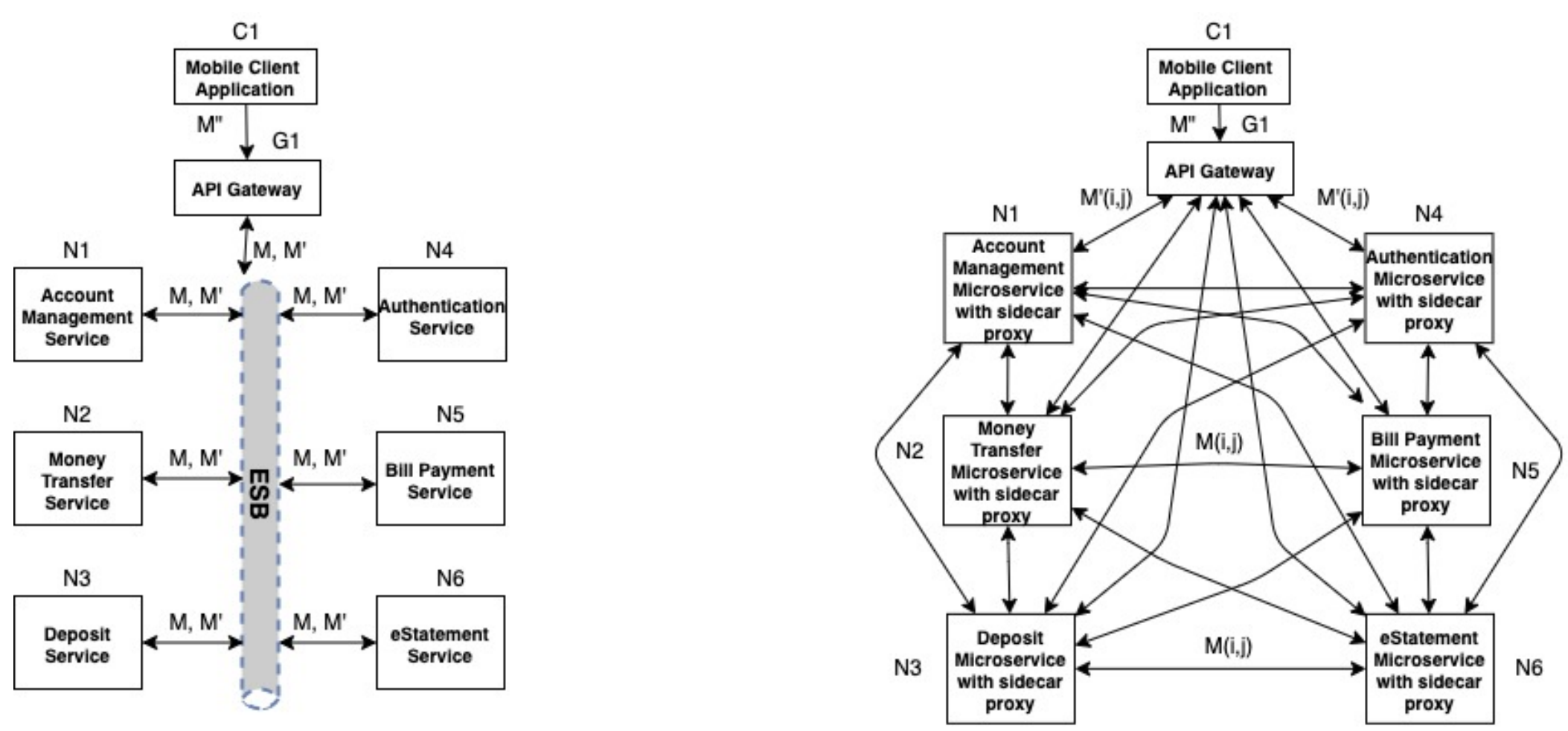

**User Stories or Use Cases:**

US1. As a customer, I want to log into my account, so that I can see my balances and account information.

US2. As a customer, I want to transfer money from my chequing account to my GIC saving account, so that I can earn some interests on my extra cash.

US3. As a customer, I want to transfer money to another individual in another bank, so that I can pay back money that I owed to the person on time.

US4. As a customer, I want to deposit a cheque into my account, so that I can use the money to pay my bills.

US5. As a customer, I want to pay my bill from my chequing account, so that I can pay my bill on time without penalty.

US6. As a customer, I want to download my eStatements, so that I can validate my payments and spendings.

**Figure 9. A Sample Retail Banking System Integration - ESB vs Service Mesh Pattern**

Having established the theoretical framework, this section demonstrates the practical application of our model through a case study of a sample retail banking integration. The objective is to evaluate two architectural patterns: Enterprise Service Bus (ESB) and Service Mesh, as illustrated in Figure 9. To instantiate the model, we align the business domain with the technical system by mapping functional requirements to technical components. This alignment is achieved by identifying the shared elements—such as nodes and edges—inherent in both domains.

Furthermore, we extend this mapping to the cost dimension by associating the technical complexity with the effort required for implementation. Specifically, we model the interaction as a set of request-response sequences, where solid lines represent outgoing requests and dotted lines represent incoming responses, as illustrated in Figure 10. While a preliminary visual inspection of the architecture suggests that the Service Mesh pattern exhibits higher complexity due to a greater density of interactions, our formal analysis provides a more nuanced comparison of the underlying cost drivers.

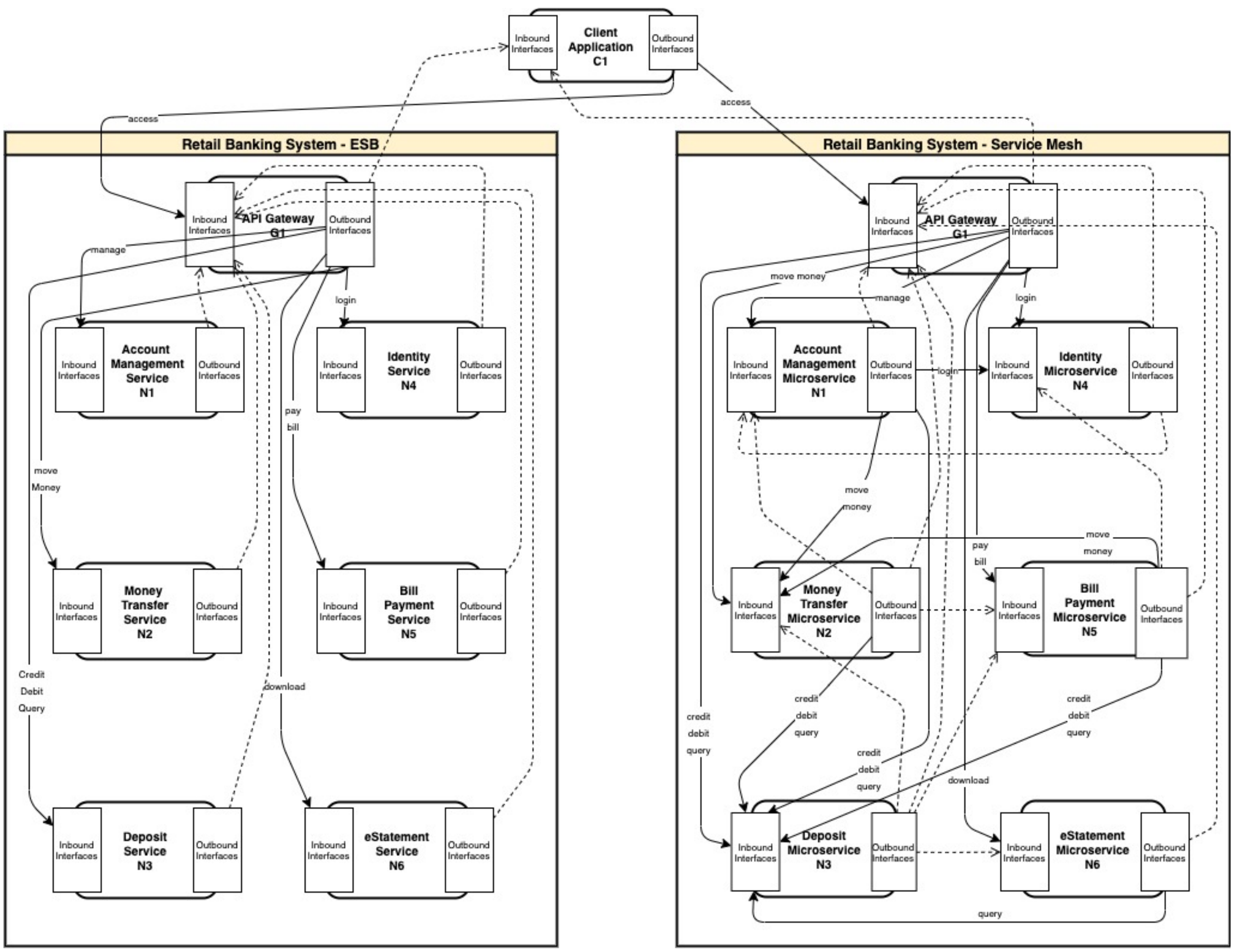

Figure 10. A CCSM for Sample Retail Banking System Integration

To facilitate practical implementation, the SCCM and SCEM models are transformed into a tabular format comprising twelve distinct calculation tables, as illustrated in Figure 11. These tables are categorized into four functional groups: Properties, Couplings, Complexity, and Cost. The structural dimensionality of these tables varies;

Property tables are one-dimensional (indexed by component), whereas Coupling tables are two-dimensional matrices representing the interactions between components.

The population of these tables relies on two distinct processes: manual assignment and algorithmic derivation. While certain intrinsic properties and cost factors are manually assigned, the remaining values are derived through the systematic application of the model's formal definitions. Specifically, the Coupling and Complexity values are computed by applying Definitions 4 through 15 to the input properties, while the final Cost tables aggregate these results to provide a complete view of the system's overhead.

**ESB - Inbound Message Property Matrix**

| | C1 | G1 | N1 | N2 | N3 | N4 | N5 | N6 | Average |
|---|---|---|---|---|---|---|---|---|---|
| Message Property | 5 | 10 | 10 | 10 | 10 | 10 | 10 | 10 | 9.4 |

**ESB - Outbound Message Property Matrix**

| | C1 | G1 | N1 | N2 | N3 | N4 | N5 | N6 | Average |
|---|---|---|---|---|---|---|---|---|---|
| Message Property | 5 | 10 | 10 | 10 | 10 | 10 | 10 | 10 | 9.4 |

**ESB - Inbound Interface Property Matrix**

| | C1 | G1 | N1 | N2 | N3 | N4 | N5 | N6 | Average |
|---|---|---|---|---|---|---|---|---|---|
| interface Property | 10 | 20 | 40 | 40 | 40 | 40 | 40 | 40 | 33.8 |

**ESB - Outbound Interface Property Matrix**

| | C1 | G1 | N1 | N2 | N3 | N4 | N5 | N6 | Average |
|---|---|---|---|---|---|---|---|---|---|
| interface Property | 10 | 20 | 20 | 20 | 20 | 20 | 20 | 20 | 18.8 |

**ESB - Inbound Interface Coupling Matrix (Create new coupling)**

| | To: C1 | G1 | N1 | N2 | N3 | N4 | N5 | N6 | Total |
|---|---|---|---|---|---|---|---|---|---|
| From: C1 | 0 | 350 | 0 | 0 | 0 | 0 | 0 | 0 | 350 |
| G1 | 350 | 0 | 400 | 400 | 400 | 400 | 400 | 400 | 2750 |
| N1 | 0 | 400 | 0 | 0 | 0 | 0 | 0 | 0 | 400 |
| N2 | 0 | 800 | 0 | 0 | 0 | 0 | 0 | 0 | 800 |
| N3 | 0 | 800 | 0 | 0 | 0 | 0 | 0 | 0 | 800 |
| N4 | 0 | 800 | 0 | 0 | 0 | 0 | 0 | 0 | 800 |
| N5 | 0 | 800 | 0 | 0 | 0 | 0 | 0 | 0 | 800 |
| N6 | 0 | 800 | 0 | 0 | 0 | 0 | 0 | 0 | 800 |
| Total | 350 | 4750 | 400 | 400 | 400 | 400 | 400 | 400 | **7500** |

**ESB - Outbound Interface Coupling Matrix (Create new coupling)**

| | To: C1 | G1 | N1 | N2 | N3 | N4 | N5 | N6 | Total |
|---|---|---|---|---|---|---|---|---|---|
| From: C1 | 0 | 350 | 0 | 0 | 0 | 0 | 0 | 0 | 350 |
| G1 | 350 | 0 | 400 | 400 | 400 | 400 | 400 | 400 | 2750 |
| N1 | 0 | 400 | 0 | 0 | 0 | 0 | 0 | 0 | 400 |
| N2 | 0 | 400 | 0 | 0 | 0 | 0 | 0 | 0 | 400 |
| N3 | 0 | 400 | 0 | 0 | 0 | 0 | 0 | 0 | 400 |
| N4 | 0 | 400 | 0 | 0 | 0 | 0 | 0 | 0 | 400 |
| N5 | 0 | 400 | 0 | 0 | 0 | 0 | 0 | 0 | 400 |
| N6 | 0 | 400 | 0 | 0 | 0 | 0 | 0 | 0 | 400 |
| Total | 350 | 2750 | 400 | 400 | 400 | 400 | 400 | 400 | **5500** |

**ESB - Internal Coupling Matrix (Create new coupling)**

| | To: C1 | G1 | N1 | N2 | N3 | N4 | N5 | N6 | Total |
|---|---|---|---|---|---|---|---|---|---|
| From: C1 | 350 | 700 | 350 | 350 | 350 | 350 | 350 | 350 | 3150 |
| G1 | 3100 | 2750 | 20 | 20 | 20 | 20 | 20 | 20 | 24750 |
| N1 | 400 | 800 | 400 | 400 | 400 | 400 | 400 | 400 | 3600 |
| N2 | 400 | 1200 | 400 | 400 | 400 | 400 | 400 | 400 | 4000 |
| N3 | 400 | 1200 | 400 | 400 | 400 | 400 | 400 | 400 | 4000 |
| N4 | 400 | 1200 | 400 | 400 | 400 | 400 | 400 | 400 | 4000 |
| N5 | 400 | 1200 | 400 | 400 | 400 | 400 | 400 | 400 | 4000 |
| N6 | 400 | 1200 | 400 | 400 | 400 | 400 | 400 | 400 | 4000 |
| Total | 5850 | 10250 | 5900 | 5900 | 5900 | 5900 | 5900 | 5900 | **51500** |

**ESB - External Coupling Matrix (Create new coupling)**

| | C1 | G1 | N1 | N2 | N3 | N4 | N5 | N6 | Total |
|---|---|---|---|---|---|---|---|---|---|
| Coupling Value | 700 | 7500 | 800 | 800 | 800 | 800 | 800 | 800 | 13000 |
| Total | 700 | 7500 | 800 | 800 | 800 | 800 | 800 | 800 | **13000** |

**ESB - Internal Coupling Cost Factor Matrix (Create new coupling in person-hour)**

| | C1 | G1 | N1 | N2 | N3 | N4 | N5 | N6 | AVG |
|---|---|---|---|---|---|---|---|---|---|
| pp-hr | 1 | 1 | 1 | 1 | 1 | 1 | 1 | 1 | 1 |
| AVG | 1 | 1 | 1 | 1 | 1 | 1 | 1 | 1 | 1 |

**ESB - External Coupling Cost Factor Matrix (Create new coupling in person-hour)**

| | C1 | G1 | N1 | N2 | N3 | N4 | N5 | N6 | AVG |
|---|---|---|---|---|---|---|---|---|---|
| pp-hr | 1 | 1 | 1 | 1 | 1 | 1 | 1 | 1 | 1 |
| AVG | 1 | 1 | 1 | 1 | 1 | 1 | 1 | 1 | 1 |

**ESB - Component Complexity Matrix (Create new coupling)**

| | C1 | G1 | N1 | N2 | N3 | N4 | N5 | N6 | Total |
|---|---|---|---|---|---|---|---|---|---|
| Interface | 350 | 4750 | 400 | 400 | 400 | 400 | 400 | 400 | 7500 |
| Impl | 6200 | 13000 | 6300 | 6300 | 6300 | 6300 | 6300 | 6300 | 57000 |
| Total | 6550 | 17750 | 6700 | 6700 | 6700 | 6700 | 6700 | 6700 | **64500** |

**ESB - Component Cost Matrix (Create new coupling in person-hour)**

| | C1 | G1 | N1 | N2 | N3 | N4 | N5 | N6 | Total |
|---|---|---|---|---|---|---|---|---|---|
| Interface | 350 | 4750 | 400 | 400 | 400 | 400 | 400 | 400 | 7500 |
| Impl | 6200 | 13000 | 6300 | 6300 | 6300 | 6300 | 6300 | 6300 | 57000 |
| Total | 6550 | 17750 | 6700 | 6700 | 6700 | 6700 | 6700 | 6700 | **64500** |

**Figure 11. Tabular Form of SCCM - A Sample ESB for Retail Banking System Integration**

For the purpose of this exercise, we also make a number of assumptions on the SCCM models for the ESB and Service Mesh integration patterns. First, the six banking services (microservice) has an average of 1 inbound and outbound interface pairs, except for the deposit service which has 3 interface pairs as well as the API gateway which has 7 interface pairs. Secondly, we assume that the property value of a message is equivalent to the magnitude of the message which is equal to the number of attributes in the message. Thirdly, the weighted constants *(INK1, INK2, INK3, INK4, OUTK1, OUTK2, OUTK3, and OUTK4)* are assigned with a value of 1. Lastly, we also assign all the cost factors *(e.g. CFELM, CFICP, CFOCP)* with an effort-cost value of 1 person-hour.

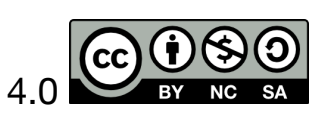

Figures 11 illustrates the tabular version of our SCCM for the ESB alternative as in Figure 10. We assume that the canonical ESB message format is twice as complex as the client-gateway message (e.g. 5 vs 10). Since it typically requires more data processing overhead on the service requests than service responses, the property value of the inbound interface on the ESB is assumed to be twice as complex as the outbound interface and four times as complex as the client-gateway interface (e.g. 10, 20, 40). The tables show interesting results. The upper bounded magnitude of complexity of the ESB alternative has a value of 64,500; 7,500 for interface (API) and 57,000 for implementation. The upper bounded cost of implementing the ESB alternative from scratch requires 64,500 person-hours and $3.87M respective; 7,500 person-hours for interface (API) and 57,000 person-hours for implementation.

**Service Mesh - Inbound Message Property Matrix**

| | C1 | G1 | N1 | N2 | N3 | N4 | N5 | N6 | Average |
|---|---|---|---|---|---|---|---|---|---|
| interface Property | 5 | 5 | 5 | 5 | 5 | 5 | 5 | 5 | 5.0 |

**Service Mesh - Outbound Message Property Matrix**

| | C1 | G1 | N1 | N2 | N3 | N4 | N5 | N6 | Average |
|---|---|---|---|---|---|---|---|---|---|
| interface Property | 5 | 5 | 5 | 5 | 5 | 5 | 5 | 5 | 5.0 |

**Service Mesh - Inbound Interface Property Matrix**

| | C1 | G1 | N1 | N2 | N3 | N4 | N5 | N6 | Average |
|---|---|---|---|---|---|---|---|---|---|
| interface Property | 10 | 10 | 5 | 5 | 5 | 5 | 5 | 5 | 6.3 |

**Service Mesh - Outbound Interface Property Matrix**

| | C1 | G1 | N1 | N2 | N3 | N4 | N5 | N6 | Average |
|---|---|---|---|---|---|---|---|---|---|
| interface Property | 10 | 10 | 5 | 5 | 5 | 5 | 5 | 5 | 6.3 |

**Service Mesh - Inbound Interface Coupling Matrix (Create new coupling)**

| | To: C1 | G1 | N1 | N2 | N3 | N4 | N5 | N6 | Total |
|---|---|---|---|---|---|---|---|---|---|
| From: C1 | 0 | 350 | 0 | 0 | 0 | 0 | 0 | 0 | 350 |
| G1 | 350 | 0 | 50 | 50 | 50 | 50 | 50 | 50 | 650 |
| N1 | 0 | 50 | 0 | 25 | 0 | 25 | 0 | 0 | 100 |
| N2 | 0 | 50 | 25 | 0 | 25 | 0 | 25 | 0 | 125 |
| N3 | 0 | 50 | 25 | 25 | 0 | 0 | 25 | 0 | 125 |
| N4 | 0 | 50 | 25 | 0 | 0 | 0 | 25 | 0 | 100 |
| N5 | 0 | 50 | 0 | 25 | 25 | 0 | 0 | 0 | 100 |
| N6 | 0 | 50 | 0 | 0 | 25 | 0 | 0 | 0 | 75 |
| Total | 350 | 650 | 125 | 125 | 125 | 75 | 125 | 50 | **1625** |

**Service Mesh - Outbound Interface Coupling Matrix (Create new coupling)**

| | To: C1 | G1 | N1 | N2 | N3 | N4 | N5 | N6 | Total |
|---|---|---|---|---|---|---|---|---|---|
| From: C1 | 0 | 350 | 0 | 0 | 0 | 0 | 0 | 0 | 350 |
| G1 | 350 | 0 | 50 | 50 | 50 | 50 | 50 | 50 | 650 |
| N1 | 0 | 50 | 0 | 25 | 0 | 25 | 0 | 0 | 100 |
| N2 | 0 | 50 | 25 | 0 | 25 | 0 | 25 | 0 | 125 |
| N3 | 0 | 50 | 25 | 25 | 0 | 0 | 25 | 0 | 125 |
| N4 | 0 | 50 | 25 | 0 | 0 | 0 | 25 | 0 | 100 |
| N5 | 0 | 50 | 0 | 25 | 25 | 0 | 0 | 0 | 100 |
| N6 | 0 | 50 | 0 | 0 | 25 | 0 | 0 | 0 | 75 |
| Total | 350 | 650 | 125 | 125 | 125 | 75 | 125 | 50 | **1625** |

**Service Mesh - Internal Coupling Matrix (Create new coupling)**

| | To: C1 | G1 | N1 | N2 | N3 | N4 | N5 | N6 | Total |
|---|---|---|---|---|---|---|---|---|---|
| From: C1 | 350 | 700 | 350 | 350 | 350 | 350 | 350 | 350 | 3150 |
| G1 | 1000 | 650 | 700 | 700 | 700 | 700 | 700 | 700 | 5850 |
| N1 | 100 | 150 | 100 | 125 | 100 | 125 | 100 | 100 | 900 |
| N2 | 125 | 175 | 150 | 125 | 150 | 125 | 150 | 125 | 1125 |
| N3 | 125 | 175 | 150 | 150 | 125 | 125 | 150 | 125 | 1125 |
| N4 | 100 | 150 | 125 | 100 | 100 | 100 | 125 | 100 | 900 |
| N5 | 100 | 150 | 100 | 125 | 125 | 100 | 100 | 100 | 900 |
| N6 | 75 | 125 | 75 | 75 | 100 | 75 | 75 | 75 | 675 |
| Total | 1975 | 2275 | 1750 | 1750 | 1750 | 1700 | 1750 | 1675 | **14625** |

**Service Mesh - External Coupling Matrix (Create new coupling)**

| | C1 | G1 | N1 | N2 | N3 | N4 | N5 | N6 | Total |
|---|---|---|---|---|---|---|---|---|---|
| Coupling Value | 700 | 1300 | 250 | 250 | 250 | 150 | 250 | 100 | 3250 |
| Total | 700 | 1300 | 250 | 250 | 250 | 150 | 250 | 100 | **3250** |

**Service Mesh - Internal Coupling Cost Factor Matrix (Create new coupling in person-hour)**

| | C1 | G1 | N1 | N2 | N3 | N4 | N5 | N6 | AVG |
|---|---|---|---|---|---|---|---|---|---|
| pp-hr | 1 | 1 | 1 | 1 | 1 | 1 | 1 | 1 | 1 |
| AVG | 1 | 1 | 1 | 1 | 1 | 1 | 1 | 1 | 1 |

**Service Mesh - External Coupling Cost Factor Matrix (Create new coupling in person-hour)**

| | C1 | G1 | N1 | N2 | N3 | N4 | N5 | N6 | AVG |
|---|---|---|---|---|---|---|---|---|---|
| pp-hr | 1 | 1 | 1 | 1 | 1 | 1 | 1 | 1 | 1 |
| AVG | 1 | 1 | 1 | 1 | 1 | 1 | 1 | 1 | 1 |

**Service Mesh - Component Complexity Matrix (Create new coupling)**

| | C1 | G1 | N1 | N2 | N3 | N4 | N5 | N6 | Total |
|---|---|---|---|---|---|---|---|---|---|
| Interface | 350 | 650 | 125 | 125 | 125 | 75 | 125 | 50 | 1625 |
| Impl | 2325 | 2925 | 1875 | 1875 | 1875 | 1775 | 1875 | 1725 | 16250 |
| Total | 2675 | 3575 | 2000 | 2000 | 2000 | 1850 | 2000 | 1775 | **17875** |

**Service Mesh - Component Cost Matrix (Create new coupling in person-hour)**

| | C1 | G1 | N1 | N2 | N3 | N4 | N5 | N6 | Total |
|---|---|---|---|---|---|---|---|---|---|
| Interface | 350 | 650 | 125 | 125 | 125 | 75 | 125 | 50 | 1625 |
| Impl | 2325 | 2925 | 1875 | 1875 | 1875 | 1775 | 1875 | 1725 | 16250 |
| Total | 2675 | 3575 | 2000 | 2000 | 2000 | 1850 | 2000 | 1775 | **17875** |

**Figure 12. Tabular Form of SCCM - A Sample Service Mesh for Retail Banking System Integration**

Figure 12 illustrates a tabular version of our SCCM for the Service Mesh alternative as in Figure 10. We assume that the microservice is lightweight; and therefore, its message format is also lightweight (e.g. 5). Since it typically requires more data processing overhead on microservice requests than microservice responses, the property value of the inbound interface on the ESB is assumed to be twice as complex as the outbound interface (e.g. 5, 10). The tables in Figure 2 show interesting results. The upper-bounded magnitude of complexity of the Service Mesh alternative has a value of 17,875; 1,625 for interface (API) and 16,250 for implementation. The upper-bounded cost of implementing the ESB alternative from scratch requires 17,875 person-hours and $1.7M respectively; 1,625 person-hours for interfaces (APIs) and 16,250 person-hours for implementation.

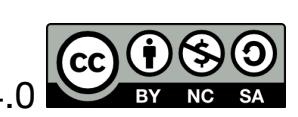

Figure 13 shows a sample architecture decision for our Retail Banking System Integration. It outlines our decisioning process by evaluating the results from the graphical and tabular forms of our formal model (Figure 10-12). Given that the build cost is the number one decision factor, the decision is to go with Service Mesh because the upper-bounded build cost is much lower than ESB ($1.07M vs $3.87M); that is almost four fold. In addition, Service Mesh appears to remain scalable and more cost-effective than ESB, provided that the microservices remain lightweight and they do not need to communicate with each other in a full mesh.

**Context, Intent, Motivations, and Applicability:**

The intent and motivation of the problem is to select an integration pattern for secure and efficient client and back-end services on a retail banking platform over the cloud.

The context is for large volume transactions in digital retail banking. It is applicable to banking customers to access their banking services over digital channels. It is also bounded to a data flow system domain. The activities in scope of the activity domain are related to constructing the integration elements and messaging flows for the chosen integration pattern.

The objectives and the driving forces are to minimize the cost of building and extending the backend legacy services while the architecture maintains the performance requirements.

Therefore, the solution needs to keep the system integration complexity to a minimum. It must also be scalable and extendible with new applications and services while it maintains the performance requirements for the future states.

**Alternative (Structure) Description:**

In the ESB pattern (as in Figure 1), a client application would initiate service requests to the API gateway as the service broker, which routes the requests to the back-end service(s).

All the back-end services communicate with each other through the API Gateway, which acts as the service broker.

In the Service Mesh pattern (as in Figure 1), a client application would initiate service requests to the API gateway, which also routes the request to the back-end microservice(s).

All the back-end services communicate directly with each other via their sidecar proxies. Internal communications among microservices bypass the API Gateway.

All messages within a service mesh may have a unique format for every point-to-point service pair, whereas all the messages in ESB have a canonical format.

**Assumptions**

1. The cost and complexity of building a new service are the most important decision factors. They are also key decision factors in the context of performance.

2. Regardless of ESB or Service Mesh, all services or microservices need to communicate and collaborate directly or indirectly with each other.

3. The cost of extending a canonical message format is much larger than a service-specific message. The hourly rate for a developer is $60USD/hr.

**Pros for ESB**

1. Fewer integration points (e.g. O(N)) for future extension as in Figures 9 and 10.

2. Easier to monitor and manage. Centralized Bus and Service Broker (API Gateway) with messages in canonical or standardized form.

2. Message Driven and Loosely Coupled.

**Pros for Service Mesh**

1. Likely cost less to build. The estimated upper-bounded cost and complexity are: $1.07M and 17,875 person-hours respectively as in Figure 12.

2. Easy to build a new integration point. Less complexity in each direct integration point. Complex Canonical Form is not required.

3. Flexible with Direct API or Message integration via Loosely Coupled Sidecar Proxies.

**Cons for ESB**

1. More costly to build. The estimated upper-bounded cost and complexity are: $3.87M and 64,500 person-hours respectively as in Figure 11.

2. A change to a canonical message usually results in extensive changes in Service Broker and minor changes in other services.

3. The Service Broker (API Gateway) can become monolithic and complex to extend in future.

**Cons for Service Mesh**

1. Harder to monitor and manage with decentralized services.

2. Likely more complex with more integration points (e.g. $O(N+N^2)$) for future extensions.

3. Possibly more costly to extend and manage if the future microservices have more communication among each other (full mesh).

**Decision and Rationale:**

1. Choose Service Mesh because the estimated upper-bounded build cost is lower than SOA ESB.

2. Service Mesh likely remains more cost effective than SOA ESB in long run. The costs factors of developing and maintaining an integration point (coupling) in a Service Mesh remain low, provided that microservices remain lightweight and they do not need to communicate with each other (full mesh).

3. The microservice are simple and efficient. It offers better performance than the services in SOA ESB.

**Figure 13. A Sample Architecture Decision for Retail Banking System Integration - ESB vs Service Mesh Pattern**

Traditional architectural analyses often restrict their scope to merely counting components and connections, neglecting critical internal systemic properties. Our proposed methodology addresses this limitation by evaluating both the inherent topology and the detailed properties of both components and the elements they comprise. This enables comprehensive scenario exploration, allowing architects to model and quantify the impact of varying or scaling fundamental constants and system properties. Crucially, we incorporate complex parameters such as inter-component coupling complexity and inherent cost factors into our decision model. We make a key differential assumption regarding integration patterns: we postulate that service mesh architectures exhibit significantly higher average interface and message property coupling values compared to traditional Enterprise Service Bus (ESB) implementations. This assumption is grounded in the established complexity and resource overhead associated with canonical message translation and processing inherent to ESBs. Furthermore, our framework is designed for practicality, demonstrating that a robust system estimation model can be constructed using a limited set of high-level input parameters, thereby enabling technologists to accurately estimate the cost and complexity profile of diverse system architectures across various defined operational scenarios.

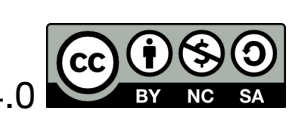

## X. Conclusion and Future Work

This paper presented a novel framework for estimating system complexity and cost by focusing on interface-driven metrics rather than internal component attributes. By utilizing the CCM, CCSM, SCCM, and SCEM models, we demonstrated that it is possible to make informed architectural decisions—such as the selection between an ESB or a Service Mesh—even when internal implementation details are unknown. Our approach moves beyond simple structural analysis to quantify the impact of external interactions, proving that an architectural decision is not merely a choice of pattern, but a calculated trade-off of cost and complexity.

The strength of this methodology lies in its versatility; the model is technology-agnostic and can be scaled from simple service interactions to highly complex, distributed ecosystems. By applying algebraic transformations to graph-based structures, we provided a way to estimate both worst-case (upper-bound) and average-case complexities, allowing architects to plan for both scalability and resource constraints. Ultimately, this work bridges the gap between high-level enterprise architecture and concrete project management, providing a mathematical foundation for the strategic deployment of technologies across diverse industries, from finance to healthcare.

While the current framework provides a robust foundation for complexity estimation, several avenues for future research remain. We intend to expand the integration of this model with automated CI/CD pipelines to enable real-time complexity monitoring during the software development lifecycle. Furthermore, we aim to refine the cost-estimation parameters by incorporating machine learning to better predict the 'cost-of-change' in evolving microservices environments. Finally, we plan to continue the development of our proprietary toolset to automate the transformation of architectural diagrams into actionable cost-complexity models, making this quantitative approach accessible to all levels of stakeholders.